\documentclass[proceedings]{JHEP3}
\usepackage{amsfonts}
\usepackage{amsmath}
\usepackage{epsfig,multicol}

\newbox\mybox

\newcommand\fverb{\setbox\mybox=\hbox\bgroup\verb}
\newcommand\fverbdo{\egroup\medskip\noindent\fbox{\unhbox\mybox}\ }
\newcommand\fverbit{\egroup\item[\fbox{\unhbox\mybox}]}
\conference{Integrable models with unstable particles}
\abstract{
We review some recent results concerning integrable quantum field theories
in 1+1 space-time dimensions which contain unstable particles in their
spectrum. Recalling first the main features of analytic scattering theories
associated to integrable models, we subsequently propose a new bootstrap
principle which allows for the construction of particle spectra involving
unstable as well as stable particles. We describe the general Lie algebraic
structure which underlies theories with unstable particles and formulate a
decoupling rule, which predicts the renormalization group flow in dependence
of the relative ordering of the resonance parameters. We extend these ideas
to theories with an infinite spectrum of unstable particles. We provide new
expressions for the scattering amplitudes in the soliton-antisoliton sector
of the elliptic sine-Gordon model in terms of infinite products of
q-deformed gamma functions. When relaxing the usual restriction on the
coupling constants, the model contains additional bound states which admit
an interpretation as breathers. For that situation we compute the complete
S-matrix of all sectors. We carry out various reductions of the model, one
of them leading to a new type of theory, namely an elliptic version of the
minimal SO(n)-affine Toda field theory.}

\title{Integrable models with unstable particles}
\author{O.A.~Castro-Alvaredo and A.~Fring\\
Institut f\"ur Theoretische Physik, Freie Universit\"at Berlin, \\
Arnimallee 14, D-14195 Berlin, Germany \\
E-mail: \email{Olalla/Fring@physik.fu-berlin.de}}

\input{tcilatex}

\begin{document}

\section{Introduction}

The structure of integrable quantum field theories (IQFT) in 1+1 space-time
dimensions has been unravelled to a very large extend. Many theories can be
solved even exactly, that is to all orders in perturbation theory, in this
context. However, the large majority of investigations concentrates on
theories which involve exclusively stable particles, despite the fact that
in nature most particles are unstable. Since of course one of the
motivations to study IQFT is to reproduce realistic features, there is an
apparent need to investigate also theories which have unstable particles in
their spectrum. The aim of this talk is to review some recent results which
deal with such theories. \footnotetext{Proceeding of the workshop on "Infinite
dimensional algebras and quantum integrable systems" 
(Faro, Portugal, July, 2003). We thank the organizers, especially Nenad
Manojlovic, for their kind hospitality and untiring engagement to make things
work.}

\section{Analytic scattering theory of factorizable integrable models}

Since not all participants of this conference work directly on integrable
quantum field theories, we briefly recall some well known facts on analytic
scattering theories in 1+1 space-time dimensions. Having in mind to
emphasize features related to unstable particles this will also be useful to
the experts. As a starting point in every scattering theory one requires a
complete set of asymptotic in and out states ($t\rightarrow \pm \infty $).
These states consist of operators $Z_{\mu }(p)$ acting on the vacuum $%
\left\vert 0\right\rangle $ and creating in this way a stable particle of
the type $\mu $ with momentum $p$. Already at this point enters the
fundamental difference between stable and unstable particles. Even though
experimentally unstable particles with a very long lifetime can very often
not be distinguished from stable ones, mathematically they are very
distinct. They can never be associated to an asymptotic state, even when
they have an extremely long lifetime, as by their very nature they will have
decayed in the infinite future or were never produced in the infinite past.
Then the scattering matrix is defined to be the operator which relates a
stable n-particle in state to a stable m-particle out state 
\begin{equation}
Z_{\mu _{m}}(\theta _{m}^{^{\prime }})\ldots Z_{\mu _{1}}(\theta
_{1}^{^{\prime }})\left\vert 0\right\rangle _{\text{out}}=S_{\mu _{1}\mu
_{2}\ldots \mu _{m}}^{\mu _{1}\mu _{2}\ldots \mu _{n}}(\theta _{1}^{^{\prime
}},\ldots \theta _{n})Z_{\mu _{1}}(\theta _{1})\ldots Z_{\mu _{n}}(\theta
_{n})\left\vert 0\right\rangle _{\text{in}}.  \label{fact}
\end{equation}%
Conveniently one parameterizes the two-momentum by the rapidity $\theta $ as 
$\vec{p}=m(\cosh \theta ,\sinh \theta )$. Now there are some very special
features happening in integrable (that means here there exists at least one
non-trivial conserved charge) quantum field theories in 1+1 dimensions \cite%
{Karowski:1977th,Zamolodchikov:1977uc,ZZ,Shankar:1978cm,Parke:1980ki}. There
is no particle production and furthermore the incoming and outgoing momenta
coincide%
\begin{equation}
\{\theta _{1}^{^{\prime }},\theta _{2}^{^{\prime }},\ldots \theta
_{m}^{^{\prime }}\}=\{\theta _{1},\theta _{2},\ldots \theta _{n}\}\qquad
\quad \text{with }n=m~.
\end{equation}%
In addition, the n-particle S-matrix factorizes into a set of two-particle
S-matrices $\!\!\!\!\!\!\!$%
\begin{equation}
S_{\mu _{1}\mu _{2}\ldots \mu _{m}}^{\mu _{1}\mu _{2}\ldots \mu _{n}}(\theta
_{1}^{^{\prime }},\ldots \theta _{n})=\!\!\!\!\!\!\prod\limits_{1\leq
i<j\leq n}\!\!\!\!\!\!S_{\mu _{i}\mu _{j}}(\theta _{i},\theta _{j})~.
\label{factt}
\end{equation}%
Obviously, this is a considerable simplification in comparison with the
general situation (\ref{fact}), as it means that once we know the
two-particle $S$-matrix, we control the entire scattering matrix. Because of
this fact, we refer from now on to the two-particle scattering matrix as 
\emph{the} $S$-matrix.

How does one construct this $S$-matrix? In general one is limited to the use
of perturbation theory in the coupling constant. In particular in higher
dimensions that is essentially the only method available. In contrast, two
dimensions are very special as they miraculously allow to determine $S$
exactly to all orders in perturbation theory. This is one of the major
successes of this area of research and one of the reasons for the continued
interest in such theories. The original ideas which lead to explicit
expressions for $S$ go back to what is called the bootstrap approach \cite%
{Schroer:1976if,Karowski:1977th,Zamolodchikov:1977uc}. It consists of using
various properties for the scattering matrix, which one motivates by some
physical principles in order to set up an axiomatic system for $S$ in the
hope that it might be so constraining that it determines $S$ completely.
Indeed, these hopes are not in vain.

We recall the $S$-matrix properties:

\noindent \underline{\textbf{i) Lorentz invariance}}

Dealing with relativistic scattering theories, we expect the scattering
matrix to be Lorentz invariant, i.e. it should depend only on covariant
combinations of the momenta. The Mandelstam variables are precisely such
quantities, see e.g. \cite{book}. In 1+1 dimensions only one of them is
independent, usually taken to be $%
s_{ab}=(p_{a}+p_{b})^{2}=m_{a}^{2}+m_{b}^{2}+2m_{a}m_{b}\cosh (\theta
_{a}-\theta _{b})$. Hence, Lorentz invariance is simply guaranteed when $S$
depends either only on $s_{ab}$ or the rapidity difference $\theta
_{ab}:=\theta _{a}-\theta _{b}$ 
\begin{equation}
S_{ab}(p_{a},p_{b})=S_{ab}(\theta _{a},\theta
_{b})=S_{ab}(s_{ab})=S_{ab}(\theta _{ab}).
\end{equation}%
Since $s_{ab}$ admits the interpretation as the total energy in the centre
of mass system, $\theta _{ab}$ has to be real for a physical process, such
that $s_{ab}\geq (m_{a}+m_{b})^{2}$.

\noindent \underline{\textbf{ii) Hermitian analyticity}}

As a central assumption of analytic $S$-matrix theory \cite{book} one
assumes that the S-matrix can be continued to the complex plane and depends
on $s_{ab},\theta _{ab}\in \mathbb{C}$. Physical scattering amplitudes are
then assumed to be real boundary values of analytic functions, which can be
obtained from a generalization of Feynman's $i\varepsilon $ prescription of
perturbation theory 
\begin{equation}
S_{ab}^{\text{physical}}=\lim_{\varepsilon \rightarrow
0}S_{ab}(s+i\varepsilon )=S_{ab}(\theta )\quad \qquad s\in \mathbb{R}%
,\varepsilon ,\theta \in \mathbb{R}^{+}.
\end{equation}%
The choice of the signs is important and relates to causality. Since a
two-particle wavefunction, having here plane waves in mind modulated by some
enveloping function, will depend on the sum of the momenta, i.e. on $\sqrt{%
s_{ab}}$, one has lost the single valuedness of the scattering matrix by an
analytic continuation. This is remedied by branch cuts along the real axis
at $s_{ab}\geq (m_{a}+m_{b})^{2}$ and $s_{ab}\leq (m_{a}-m_{b})^{2}$, the
latter being motivated by crossing see iv). Hermitian analyticity is now a
postulate which states how to continue over these cuts \cite%
{David,Miramontes:1999gd} 
\begin{equation}
\lim_{\varepsilon \rightarrow 0}S_{ab}(s+i\varepsilon )=\lim_{\varepsilon
\rightarrow 0}S_{ab}(s-i\varepsilon )\quad \Leftrightarrow \quad
S_{ab}(\theta )=\left[ S_{ba}(-\theta ^{\ast })\right] ^{\ast }~.  \label{HA}
\end{equation}%
once more for $s\in \mathbb{R},\varepsilon ,\theta \in \mathbb{R}^{+}$. The
equivalence is due to the fact that the analytic continuation $%
s+i\varepsilon \leftrightarrow s-i\varepsilon $ corresponds to $\theta
\leftrightarrow -\theta $. Often one merely uses real analyticity $%
S_{ab}(\theta )=\left[ S_{ab}(-\theta ^{\ast })\right] ^{\ast }$ instead of (%
\ref{HA}), which only coincides when $S_{ab}=S_{ba}$, that is in parity
invariant theories. This difference is very important with regard to the
theories consider below, which involve unstable particles as they
unavoidably break parity invariance. Further support for (\ref{HA}) comes
from perturbation theory \cite{David}, general considerations in analytic
S-matrix theory \cite{David2,book} and explicitly constructed examples.

\noindent \underline{\textbf{iii) Unitarity}}

Assuming that the states in (\ref{fact}) are complete and orthogonal, the
operator which maps them to each other has to be unitary 
\begin{equation}
SS^{\dagger }=S^{\dagger }S=1~.  \label{Uni}
\end{equation}%
The combination of (\ref{HA}) and (\ref{Uni}) leads to the simpler relation $%
S_{ab}(\theta )S_{ba}(-\theta )=1$, which may also be derived from applying
twice the Zamolodchikov algebra $Z_{a}(\theta _{1})Z_{b}(\theta
_{2})=S_{ab}(\theta _{12})Z_{b}(\theta _{2})Z_{a}(\theta _{1})$.

\noindent \underline{\textbf{iv) Crossing symmetry}}

Crossing symmetry can be motivated by the Lehmann-Symanzik-Zimmermann (LSZ)
formalism \cite{Lehmann:1955rq} and consists of the replacement of an
incoming particle $a$ by its anti-particle $\bar{a}$ with reversed momentum.
A discussion of the anti-particle theorem can be found in \cite{David2}. The
prescription amounts to the continuation of the Mandelstam variable $s_{ab}$
to the variable $t_{ab}=(p_{a}-p_{b})^{2}$ 
\begin{equation}
\lim_{\varepsilon \rightarrow 0}S_{ab}(s+i\varepsilon )=\lim_{\varepsilon
\rightarrow 0}S_{b\bar{a}}(t-i\varepsilon )\quad \Leftrightarrow \quad S_{b%
\bar{a}}(\theta )=S_{ab}(i\pi -\theta )~.
\end{equation}%
It is easy to check that the analytic continuation $s+i\varepsilon
\leftrightarrow t-i\varepsilon $ corresponds to $\theta \leftrightarrow i\pi
-\theta $.

\noindent \underline{\textbf{v) Yang-Baxter equation}}

In (\ref{factt}) we already indicated that the conserved charge(s) of an
integrable theory can be used to disentangle an n-particle scattering
process into a consecutive scattering of two particles only. An additional
consequence of this argument is that the order in which this takes place
does not matter, such that two different orderings are taken to be
equivalent. As in general the S-matrices do not commute, this leads to a new
constraint. In other words this amounts to say that the operators $Z$ in (%
\ref{fact}) obey an associative algebra. As a result of this one obtains the
Yang-Baxter equation \cite{Yang,Baxter} 
\begin{equation}
S(\theta _{12})\otimes S(\theta _{13})\otimes S(\theta _{23})=S(\theta
_{23})\otimes S(\theta _{13})\otimes S(\theta _{12})~.  \label{YB}
\end{equation}%
For diagonal theories, i.e. when backscattering is absent, we simply have $%
S_{ab}^{cd}(\theta )\rightarrow S_{ab}(\theta )$ such that (\ref{YB}) is
trivially satisfied.

\noindent \underline{\textbf{vi) Fusing bootstrap equation}}

By the same reasoning as in v) integrability, i.e. factorizability, yields a
further constraining equation, when two particles are allowed (what that
means is discussed in vii)) to form bound state. For instance, the particles 
$a,b$ fuse to a third particle $\bar{c}$, i.e. $a+b\rightarrow \bar{c}$. One
makes now a further assumption, sometimes referred to as nuclear democracy,
namely that also the particle of type $\bar{c}$ exists asymptotically. Then,
by integrability, it is equivalence if a third particle, say $l$, scatters
with the bound state $\bar{c}$ or consecutively with the two particles $a,b$%
. For $S$ this reads 
\begin{equation}
S_{l\bar{c}}(\theta )=S_{la}(\theta +i\bar{\eta}_{ac}^{b})S_{lb}(\theta -i%
\bar{\eta}_{bc}^{a})~.  \label{boot}
\end{equation}%
The $\bar{\eta}_{ac}^{b}\in \mathbb{R}^{+}$ are the fusing angles specific
to the individual theory considered. It is clear that the assumption of
nuclear democracy does not hold if $\bar{c}$ is an unstable particle, such
that (\ref{boot}) can not be valid in the form stated for that case. We will
now indicate the origin for the possibility to form bound states, which is
the

\noindent \underline{\textbf{vii) Pole structure}}

In general, the $S$-matrix can have a quite intricate singularity structure
consisting of poles of finite order distributed all over the complex $%
s,\theta $-plane. A strong further constraint is to assume that all
singularities which emerge in $S$ admit a consistent explanation. As a
slightly weaker assumption one could suppose that all explainable poles form
a coherent system, in the sense that the bootstrap (\ref{boot}) closes etc.,
and allow some redundant poles.

Single order poles are most important as they determine the particle
spectrum of the theory. In the $s$-plane they might be on the real axis
between the two branch cuts at $s=m_{\bar{c}}^{2}$, interpreted as an
on-shell bound state particle, or in the second Riemann sheet at $s=(m_{\bar{%
c}}-i\Gamma _{\bar{c}}/2)^{2}$ corresponding to an unstable particle with
finite lifetime $\tau =1/\Gamma _{\bar{c}}$. The discussion is more
conveniently carried out in the $\theta $-plane, since $S(\theta )$ is a
meromorphic function unlike $S(s)$. Near the singularity $S$ has to be of
the form%
\begin{equation}
S_{ab}(\theta )\sim \frac{iR_{ab}^{c}}{(\theta -i\eta _{ab}^{c}+\sigma
_{ab}^{c})}~.  \label{sing}
\end{equation}%
Depending on the location and signs of the residues we have the following
interpretations 
\begin{equation*}
\begin{array}{ll}
\text{s-channel bound state:} & R_{ab}^{c}\in \mathbb{R}^{+},\eta
_{ab}^{c}\in \mathbb{R}^{+},\sigma _{ab}^{c}=0 \\ 
\text{t-channel bound state:} & R_{ab}^{c}\in \mathbb{R}^{-},\eta
_{ab}^{c}\in \mathbb{R}^{+},\sigma _{ab}^{c}=0 \\ 
\text{unstable particle:} & R_{ab}^{c}\in \mathbb{R}^{{}},\eta _{ab}^{c}\in 
\mathbb{R}^{-},\sigma _{ab}^{c}\in \mathbb{R}^{-}%
\end{array}%
\end{equation*}%
The relation between the poles in the $s$ and $\theta $ planes are the
Breit-Wigner (BW) equations \cite{BW} 
\begin{eqnarray}
m_{_{\bar{c}}}^{2}{}-\frac{\Gamma _{_{\bar{c}}}^{2}{}}{4}
&=&m_{a}^{2}{}+m_{b}^{2}{}+2m_{a}m_{b}\cosh \sigma _{ab}^{\bar{c}}\cos \eta
_{ab}^{\bar{c}}  \label{BW1} \\
m_{_{\bar{c}}}\Gamma _{_{\bar{c}}} &=&2m_{a}m_{b}\sinh \sigma _{ab}^{\bar{c}%
}\sin \eta _{ab}^{\bar{c}}\,\,,  \label{BW2}
\end{eqnarray}%
which allow to express the mass $m_{_{_{\bar{c}}}}$ and decay width $\Gamma
_{_{\bar{c}}}$ of the unstable particle as functions of $m_{a},m_{b},\eta
_{ab}^{\bar{c}},\sigma _{ab}^{\bar{c}}$. For the stable particle formation
we have the following relation between the fusing angles in (\ref{boot}) and
the poles in (\ref{sing}): $\bar{\eta}_{ac}^{b}=\pi -\eta _{ab}^{c}$. Note
further that for the unstable particle formation in (\ref{sing}) we made the
definite choice that the unstable particle $\bar{c}$ is formed in the process%
\begin{equation}
a+b\rightarrow \bar{c}  \label{ab}
\end{equation}%
rather than in the not equivalent one $b+a$. It is clear that parity has to
be broken, as with the choice $\eta _{ab}^{c},\sigma _{ab}^{c}\in \mathbb{R}%
^{-}$ the amplitude $S_{b\bar{a}}(\theta )$ will have a pole at $i\pi -i\eta
_{ab}^{c}+\sigma _{ab}^{c}$, leaving in (\ref{BW2}) the choice that either $%
m_{_{\bar{c}}}<0$ or $\Gamma _{_{\bar{c}}}<0$, which is of course both
non-physical.

Below we will be particularly interested in the situation for large
resonance parameters $\sigma _{ab}^{\bar{c}}$, when the mass of the unstable
particles can be approximated as 
\begin{equation}
m_{\bar{c}}\sim \sqrt{m_{a}m_{b}}e^{-\sigma _{ab}^{\bar{c}}/2}\,.  \label{m}
\end{equation}

In terms of perturbation theory in a coupling constant $\beta $, the pole (%
\ref{sing}) would be of second order, i.e. $R_{ab}^{c}(\beta ^{2})$,
corresponding to a tree diagram. Similarly, higher order poles admit
interpretations in form of more complicated singular Feynman diagrams. In
some simple theories, such as for example sine-Gordon, the highest order of
the poles is two. In that context it was suggested \cite{CT} that such type
of poles are of order $\beta ^{4}$ and may be viewed as box diagrams. For
quite some time higher order poles were ignored and also here we will not
enter into a deeper discussion of them, which can be found for instance in 
\cite{Braden:1990bu,CM}. From what is said it is clear that such poles will
not alter the particle spectrum. Nonetheless, one should be able to draw the
relevant Feynman diagrams, which means one needs certain three-point
couplings to be non-vanishing. Consequently this is a constraint on the
existence of certain three point couplings $R_{ab}^{c}$.

Remarkably, \emph{the constraints i)-vii) allow to determine the }$S$\emph{%
-matrix exactly, that is to all orders in perturbation theory. }However, one
should say that the solution constructed this way is not unique, as there
exists always the possibility to multiply with so-called CDD-factors \cite%
{CDD}. To fix them requires additional arguments beyond the scheme outlined
above, such as ultraviolet limits, certain inputs from Lagrangians, etc.

\subsection{A proposal for a construction principle of unstable particle
spectra}

We have seen in the previous section, that there exists a powerful
construction principle for the spectrum of stable particles, consisting of
solving the equations (axioms) i)-vii). For unstable particles we do not
have yet such a construction tool, as by now they emerge rather passively as
poles in the unphysical sheet as by-products in the scattering process of
two stable particles. Furthermore, a description of the scattering process
of an unstable particle with another stable or unstable particle is entirely
missing in this context. Obviously, scattering processes involving unstable
particles do occur in nature, such that the quest for a proper prescription
is of physical relevance. In addition, one aims of course always at a
description which has predictive power.

From what has been said, it is clear that such a description can not be a
scattering theory in the usual sense, since for that one requires the
particles involved to exist asymptotically. Any unstable particle will
vanish in this limit rendering such formulation meaningless at first sight.
Nonetheless, some particles have extremely long lifetimes, and seem to exist
quasi infinitely long from an experimentalists point of view. It appears
therefore natural to seek a principle closely related to the conventional
bootstrap for stable particles. Inspired by this we proposed \cite{C30F} the
following construction principle:

Let us assume that in the time interval $0<t<\tau _{\bar{c}}$ we can
formally associate to the unstable particle some creation operator $\tilde{Z}%
_{\bar{c}}^{\dagger }(\theta )$, with $\lim_{t\rightarrow \infty }\tilde{Z}_{%
\bar{c}}^{\dagger }(\theta )=1$ if $\tau _{\bar{c}}<\infty $. It is clear
that these operators do not exist asymptotically, but for the stated time
interval they can mimic an asymptotic state. Let us now further suppose that
these operators satisfy a Zamolodchikov algebra%
\begin{eqnarray}
Z_{a}(\theta _{1})\tilde{Z}_{b}(\theta _{2}) &=&\tilde{S}_{ab}(\theta _{12})%
\tilde{Z}_{b}(\theta _{2})Z_{a}(\theta _{1})  \label{Z1} \\
\tilde{Z}_{a}(\theta _{1})\tilde{Z}_{b}(\theta _{2}) &=&\tilde{S}%
_{ab}(\theta _{12})\tilde{Z}_{b}(\theta _{2})\tilde{Z}_{a}(\theta _{1})
\label{Z2}
\end{eqnarray}%
which can be used to generate an S-matrix type of amplitude $\tilde{S}_{ab}$%
, describing the scattering of one unstable particle with a stable one (\ref%
{Z1}) or the scattering of two unstable particles (\ref{Z2}). We may proceed
as before and ask which type of properties might be satisfied for $\tilde{S}$%
.

\noindent \underline{\textbf{i) Lorentz invariance}}

As already indicated in (\ref{Z1}), (\ref{Z2}) it is natural to expect
Lorentz invariance also for this amplitude such that $\tilde{S}$ depends
only the rapidity difference $\theta _{ab}$ 
\begin{equation}
\tilde{S}_{ab}(p_{a},p_{b})=\tilde{S}_{ab}(\theta _{a},\theta _{b})=\tilde{S}%
_{ab}(s_{ab})=\tilde{S}_{ab}(\theta _{ab}).
\end{equation}

\noindent \underline{\textbf{ii,iii) Hermitian analyticity, unitarity}}

We will not make any assumption on hermitian analyticity here and in fact we
do not expect unitarity to hold, since the states formed with the $\tilde{Z}$
are not complete. However, applying (\ref{Z1}) or (\ref{Z2}) twice yields 
\begin{equation}
\tilde{S}_{ab}(\theta )\tilde{S}_{ba}(-\theta )=1,  \label{st1}
\end{equation}%
which also holds for $S$, derivable from combining (\ref{HA}) and (\ref{Uni}%
) in that case. In fact, also for the construction of $S$ it is really only
the corresponding equation \ to (\ref{st1}) which is employed, rather than
individually (\ref{HA}) and (\ref{Uni}).

\noindent \underline{\textbf{iv) Crossing symmetry}}

The validity of crossing can also be argued as before, but now we have to
continue as 
\begin{equation}
\lim_{\varepsilon \rightarrow 0}\tilde{S}_{ab}(s-i\varepsilon
)=\lim_{\varepsilon \rightarrow 0}\tilde{S}_{b\bar{a}}(t+i\varepsilon )\quad
\Leftrightarrow \quad \tilde{S}_{b\bar{a}}(-\theta )=\tilde{S}_{ab}(i\pi
+\theta )~,  \label{st2}
\end{equation}%
which in the $\theta $-plane amounts to the same equation as the one for $S$.

\noindent \underline{\textbf{v) Yang-Baxter equation}}

Supposing the algebra related to (\ref{Z1}), (\ref{Z2}) is associative we
have by the same reasoning as for stable particles the Yang-Baxter equation 
\begin{equation}
\tilde{S}(\theta _{12})\otimes \tilde{S}(\theta _{13})\otimes \tilde{S}%
(\theta _{23})=\tilde{S}(\theta _{23})\otimes \tilde{S}(\theta _{13})\otimes 
\tilde{S}(\theta _{12})~.
\end{equation}

\noindent \underline{\textbf{vi) Fusing bootstrap equation}}

We commence with the fusing of two stable particles to create an unstable
particle as in the process (\ref{ab}). To this process we can associate
bootstrap equations almost in the usual way. We scatter for this with an
additional stable or unstable particle, say of type $l$, \ and obtain the $%
\tilde{S}$ bootstrap equations 
\begin{equation}
\tilde{S}_{la}(\theta -\bar{\gamma}_{ca}^{\bar{b}})\,\tilde{S}_{lb}(\theta +%
\bar{\gamma}_{bc}^{\bar{a}})\,=\tilde{S}_{l\bar{c}}(\theta ),  \label{b2}
\end{equation}%
where $\bar{\gamma}=\pm \QTR{sl}{i}\pi -$ $\gamma $, $\gamma =i\eta -\sigma $
and also $\bar{\gamma}\rightarrow -\bar{\gamma}$ is not a symmetry. The
angles should be measured anti-clockwise, which explains the signs. We also
note that we do not assume parity invariance, such that in general $\bar{%
\gamma}_{ba}^{\bar{c}}\neq \bar{\gamma}_{ab}^{\bar{c}}$. With the help of (%
\ref{st1}), (\ref{st2}) one derives the bootstrap equations for the opposite
parity and the ones for the crossed processes $a+c\rightarrow \bar{b}$ and $%
b+c\rightarrow \bar{a}$ \ and from (\ref{b2}) 
\begin{eqnarray}
\tilde{S}_{\bar{c}l}(\theta ) &=&\tilde{S}_{al}(\theta +\bar{\gamma}_{ca}^{%
\bar{b}})\,\tilde{S}_{bl}(\theta -\bar{\gamma}_{bc}^{\bar{a}})\,,
\label{b33} \\
\tilde{S}_{l\bar{\jmath}}(\theta ) &=&\,\tilde{S}_{lc}(\theta -\bar{\gamma}%
_{bc}^{\bar{a}})\,\tilde{S}_{la}(\theta \pm \QTR{sl}{i}\pi -\bar{\gamma}%
_{ca}^{\bar{b}}-\bar{\gamma}_{bc}^{\bar{a}}),  \label{b3} \\
\tilde{S}_{l\bar{\imath}}(\theta ) &=&\tilde{S}_{lc}(\theta +\bar{\gamma}%
_{ca}^{\bar{b}})\tilde{S}_{lb}(\theta \pm \QTR{sl}{i}\pi +\bar{\gamma}_{ca}^{%
\bar{b}}+\bar{\gamma}_{bc}^{\bar{a}})\,\,.  \label{b4}
\end{eqnarray}%
From the crossing relation for the \textquotedblleft scattering matrix" and (%
\ref{b3}) or (\ref{b4}) one obtains some relations between the various
fusing angles 
\begin{equation}
\bar{\gamma}_{ab}^{\bar{c}}+\bar{\gamma}_{ca}^{\bar{b}}+\bar{\gamma}_{bc}^{%
\bar{a}}=\pm \QTR{sl}{i}\pi \,\,.  \label{eta}
\end{equation}%
At first sight this looks very much like the usual bootstrap prescription,
but there are some differences. As is clear from the scattering process of
two stable particles producing an unstable one, the angle $\bar{\gamma}%
_{ab}^{\bar{c}}$ is not purely complex any longer as it is for the situation
when exclusively stable particles scatter. As a consequence, this property
then extends to the other angles $\bar{\gamma}_{ca}^{\bar{b}}$ and $\bar{%
\gamma}_{bc}^{\bar{a}}$ in (\ref{b2}), which also possess some non-vanishing
real parts. Note that (\ref{eta}) implies that the real parts of the three
angels involved add up to zero. At this point we do not have an entirely
compelling reason for demanding that, but this formulation will turn out to
work well.

Of course the above equations are only a proposal, which needs to be put on
more solid ground. Nonetheless, at this point our proposal gains support
from self-consistency and its predictive power, which may be double checked:
a) The bootstrap closes consistently for many non-trivial examples, which we
calculated. As for stable particles this is never guaranteed and by no means
self-evident. b) The bootstrap yields the amount of unstable particles
together with their mass. This prediction can be used to explain a mass
degeneracy of some unstable particles which can not be seen in a
thermodynamic Bethe ansatz (TBA) analysis for the concrete example of the
homogeneous sine-Gordon (HSG) models, see below. c) The bootstrap is in
agreement with a general Lie algebraic decoupling rule, which we also
present below, describing the behaviour when certain resonance parameters
tend to infinity. d) The bootstrap yields the three-point couplings of all
possible interactions, that is, involving stable as well as unstable
particles.

\subsection{An example: The g$_{k}$-HSG model}

The $\mathbf{g}_{k}$-homogeneous sine-Gordon models (HSG) \cite%
{Park:1994bx,Fernandez-Pousa:1997hi}, with $\mathbf{g}$\textbf{\ }being a
simple Lie algebra of rank $\ell $ and level $k$, will be our standard
example in what follows. In fact they have been the first models with a well
defined Lagrangian containing unstable particles which have been the subject
of a systematic analysis \cite%
{SHSG,Castro-Alvaredo:1999em,CFK,Ident,CF8,Ren,Miramontes:2000wt,C30F,Dorey:2002sc}%
. They can be viewed as perturbed conformal field theories (CFTs)\footnote{%
For the particular case of the SU(3)$_{2}$-HSG model it was shown \cite{Pas}
that it can be described alternatively as a perturbation of a tensor product
of two minimal CFTs.} 
\begin{equation}
\mathcal{H}_{G_{k}\text{-HSG}}=\mathcal{H}_{G_{k}/U(1)^{\ell }\text{-CFT}%
}-\lambda \int d^{2}x\phi (x,t)\,.  \label{pert}
\end{equation}%
The underlying ultraviolet CFT is a Wess-Zumino-Novikov-Witten-$%
G_{k}/U(1)^{\ell }$-coset theory \cite{Witten,GKO}. The correspondig
Virasoro central charge $c\,\ $is computed with standard arguments of \cite%
{GKO}$\ $and the perturbing operator $\phi $ is identified with a primary
field of conformal dimensions $\Delta ,\bar{\Delta}$. One finds 
\begin{equation}
c=\ell \,\frac{k\,h-h^{\vee }}{k+h^{\vee }}\qquad \text{and\qquad }\Delta =%
\bar{\Delta}=\frac{h^{\vee }}{k+h^{\vee }}\,.  \label{cdel}
\end{equation}%
Here $(h^{\vee })\,h$ is the (dual) Coxeter number of $\mathbf{g}$. For
simplicity we will drop in the following the explicit mentioning of the
subalgebra $U(1)^{\ell }$ which were indicated in (\ref{pert}).

The scattering matrix for $\mathbf{g}_{k}$-HSG-models with $\mathbf{g}$
simply laced algebras was constructed in \cite{SHSG}. For $k=2$ it can be
brought into the simple form 
\begin{equation}
S_{ij}(\theta ,\sigma _{ij})=(-1)^{\delta _{ij}}\,\varepsilon (\sigma
_{ij})(\sigma _{ij},2)^{I_{ij}}\,,\qquad 1\leq i,j\leq \ell  \label{SHSG}
\end{equation}%
where $I$ denotes the incidence matrix of $\mathbf{g}$\textbf{\ }and $%
\varepsilon (x)$ is the step-function, i.e.~$\varepsilon (x)=1$ for $x\geq 0$%
, $\varepsilon (x)=-1$ for $x<0$.\textbf{\ }It is convenient to use the
abbreviation 
\begin{equation}
(\sigma ,x):=\tanh (\theta +\sigma -i\pi x/4)/2\,.  \label{building}
\end{equation}

Let us now consider the concrete case $SU(3)_{2}$. We can start with the
known part of the scattering matrix (\ref{SHSG}) for the stable particles,
and leave the remaining entries which involve unstable particles unknown.
From this we construct consistent solutions to the bootstrap equations (\ref%
{b2}), (\ref{b3}) and (\ref{b4}). We can fix the imaginary parts of the
fusing angles by the requirement that for vanishing resonance parameters we
want to reproduce the masses predicted by the Breit-Wigner formula. Choosing
the masses of the stable particles to be $m_{1}=m_{2}=m$ the one for the
unstable results to $m_{(12)}=\sqrt{2}m$. This argument does not constrain
the real parts of the fusing angles, such that they are not completely fixed
and still contain a certain ambiguity. The different choices of these
parameters give rise to slightly different theories. First we consider the
case $\sigma _{21}>0$.

\unitlength=1.0cm 
\begin{picture}(14.20,1.4)(-4.2,-0.70)
\put(0,0.00){\circle*{0.2}}
\put(-0.25,-0.50){$\alpha_1$}
\put(0.00,-0.01){\line(1,0){0.9}}
\put(0.35,-0.01){\line(2,1){0.4}}
\put(0.35,-0.01){\line(2,-1){0.4}}
\put(1.00,0.00){\circle*{0.2}}
\put(1.0,-0.50){$\alpha_2$}
\end{picture}

\noindent For this choice of the resonance parameter, we then find the
following bootstrap equations 
\begin{equation}
\tilde{S}_{l(12)}(\theta )=\tilde{S}_{l1}(\theta +(1-\nu )\sigma _{12}+i\pi
/4)\tilde{S}_{l2}(\theta -\nu \sigma _{12}-i\pi /4)
\end{equation}%
from which we construct 
\begin{equation}
\tilde{S}_{SU(3)}(\theta ,\sigma _{12})=\left( 
\begin{array}{ccc}
-1 & -(\sigma _{12},2) & -((1-\nu )\sigma _{12},3) \\ 
(\sigma _{21},2) & -1 & -(\nu \sigma _{21},1) \\ 
-((\nu -1)\sigma _{12},1) & -(\nu \sigma _{12},3) & -1%
\end{array}%
\right) \,.  \label{SU3}
\end{equation}%
Here we label the rows and columns in the order $\left\{ 1,2,(12)\right\} $.
According to the principles outlined above, the \~{S}-matrix (\ref{SU3})
allows for the processes 
\begin{equation}
1+2\rightarrow (12),\qquad 2+(12)\rightarrow 1,\qquad (12)+1\rightarrow 2.\,
\label{pp}
\end{equation}%
The related fusing angles are read off from (\ref{SU3}) as 
\begin{equation}
\gamma _{12}^{(12)}=-\frac{i\pi }{2}+\sigma _{21},\quad \gamma _{(12)1}^{2}=-%
\frac{3i\pi }{4}+(1-\nu )\sigma _{12},\quad \gamma _{2(12)}^{1}=-\frac{3i\pi 
}{4}+\nu \sigma _{12}\,
\end{equation}%
and are interrelated through equation (\ref{eta}), which still holds even
though the $\gamma $'s have non-vanishing real parts. We can employ these
fusing angles and compute the masses and decay widths by means of the
Breit-Wigner formulae (\ref{BW1}) and (\ref{BW2}). Taking again for
simplicity $m_{1}=m_{2}=m$ and in addition $\nu =1/2$, we obtain for the
first process in (\ref{pp}) 
\begin{equation}
m_{(12)}=\sqrt{2}m\cosh \sigma _{21}/2\quad \text{and\quad }\Gamma _{(12)}=2%
\sqrt{2}m\sinh \sigma _{21}/2\,.
\end{equation}%
Employing now also in the process $2+(12)\rightarrow 1$ the Breit-Wigner
formula, we reproduce in the limit $\sigma _{12}\rightarrow 0$ the values $%
m_{1}=m$ and $\Gamma _{1}=0$. Likewise, in the last process in (\ref{pp}) we
obtain $m_{2}=m$ and $\Gamma _{2}=0$.

The asymptotic limit $t\rightarrow \infty $ becomes meaningful when we
operate on an energy scale at which the unstable particle has not even been
created yet, i.e.~$\Gamma _{(12)}\rightarrow \infty \equiv \sigma
_{21}\rightarrow \infty $. In that case the theory decouples into two SU(2)$%
_{2}$-models, i.e.~free Fermions, with $S_{11}=S_{22}=-1$. This is a simple
version of the decoupling rule (\ref{drule}).

Next we consider a different theory with $\sigma _{12}>0$.

\unitlength=1.0cm 
\begin{picture}(14.20,1.4)(-4.2,-.70)
\put(0,0.00){\circle*{0.2}}
\put(-0.25,-0.50){$\alpha_1$}
\put(0.00,-0.01){\line(1,0){0.9}}
\put(0.6,-0.01){\line(-2,1){0.4}}
\put(0.6,-0.01){\line(-2,-1){0.4}}
\put(1.00,0.00){\circle*{0.2}}
\put(1.0,-0.50){$\alpha_2$}
\end{picture}

\noindent Taking also in this case for simplicity $\nu =1/2$, we find the
following bootstrap satisfied 
\begin{equation}
\tilde{S}_{l(12)}(\theta )=\tilde{S}_{l2}(\theta -\sigma _{12}/2+i\pi /4)%
\tilde{S}_{l1}(\theta +\sigma _{12}/2-i\pi /4),
\end{equation}%
which yields the S-matrix 
\begin{equation}
\tilde{S}_{SU(3)}(\theta ,\sigma _{21})=\left( 
\begin{array}{ccc}
-1 & (\sigma _{12},2) & -(\sigma _{12}/2,1) \\ 
-(\sigma _{21},2) & -1 & -(\sigma _{21}/2,3) \\ 
-(\sigma _{21}/2,3) & -(\sigma _{12}/2,1) & -1%
\end{array}%
\right) \,.  \label{SU32}
\end{equation}%
The S-matrix (\ref{SU32}) allows for the processes 
\begin{equation}
2+1\rightarrow (12),\qquad 1+(12)\rightarrow 2,\qquad (12)+2\rightarrow 1,\,
\end{equation}%
instead of (\ref{pp}). Now the fusing angles are read off as 
\begin{equation}
\gamma _{21}^{(12)}=-\frac{i\pi }{2}+\sigma _{12},\quad \gamma _{1(12)}^{2}=-%
\frac{3i\pi }{4}-\frac{\sigma _{12}}{2},\quad \gamma _{2(12)}^{1}=-\frac{%
3i\pi }{4}-\frac{\sigma _{12}}{2}\,
\end{equation}%
and also satisfy (\ref{eta}). The masses and decay width are obtained again
from (\ref{BW1}) and (\ref{BW2}) with $\sigma _{12}\rightarrow \sigma _{21}$%
. As a whole, we can think of this theory simply as being obtained from the $%
\mathbb{Z}_{2}$-Dynkin diagram automorphism which exchanges the roles of the
particles $1$ and $2$. However, since parity invariance is now broken this
is not a symmetry any more and the two theories are different. In the
asymptotic limit $\sigma _{12}\rightarrow \infty $, we obtain once again a
simple version of the decoupling rule (\ref{drule}) and the theory decouples
into two SU(2)$_{2}$-models.

The next example, SU(4)$_{2}$-HSG, is more intriguing as it leads to the
prediction a new unstable particle. Proceeding in the way as before we
construct the corresponding amplitudes $\tilde{S}$, for details see \cite%
{C30F}. We found there the processes 
\begin{equation}
\begin{array}{rrr}
1+2\rightarrow \,\,(12),\quad & (12)+1\rightarrow \quad \,2,\quad & 
2+(12)\rightarrow \quad \,1, \\ 
3+2\rightarrow \,\,(23),\quad & (23)+3\rightarrow \quad \,2,\quad & 
2+(23)\rightarrow \quad \,3,%
\end{array}
\label{27}
\end{equation}%
which simply correspond to two copies of SU(3)$_{2}$-HSG. It is interesting
to note that the amplitudes $\tilde{S}_{(12)3}$ and $\tilde{S}_{(23)1}$
contain poles at 
\begin{equation}
\gamma _{(12)3}^{(123)}=\frac{\sigma _{21}-2\sigma _{23}}{2}-\frac{3i\pi }{4}%
\quad \text{and\quad }\gamma _{(23)1}^{(123)}=\frac{\sigma _{23}-2\sigma
_{21}}{2}-\frac{3i\pi }{4}~,
\end{equation}%
which yield the possible processes 
\begin{equation}
\begin{array}{rrr}
(12)+3\rightarrow (123),\quad & (123)+(12)\rightarrow \quad 3,\quad & 
3+(123)\rightarrow (12), \\ 
(23)+1\rightarrow (123),\quad & (123)+(23)\rightarrow \quad \,1,\quad & 
1+(123)\rightarrow (23).%
\end{array}
\label{28}
\end{equation}%
An interesting prediction results from the consideration of the first two
processes in (\ref{28}). Making in the first process the particle $(12)$ and
in the second the particle $(23)$ stable, by $\sigma _{2}\rightarrow \sigma
_{1}$ and by $\sigma _{2}\rightarrow \sigma _{3}$, respectively, both
predict the mass of the particle $(123)$ as 
\begin{equation}
m_{(123)}\sim me^{|\sigma _{13}|/2}\;.  \label{m13}
\end{equation}%
This value is precisely the one we expect from the approximation in the
Breit-Wigner formula (\ref{m}). Note that in one case we obtain $\sigma
_{13} $ and in the other $\sigma _{31}$ as a resonance parameter. The
difference results from the fact that according to the processes (\ref{28}),
the particle $(123)$ is either formed as $(1+2)+3$ or $3+(2+1)$. Thus the
different parity shows up in this process, but this has no effect on the
values for the mass.

In \cite{C30F} we presented more examples and remarkably we found
consistency in each case. We take the closure of the bootstrap equations as
a non-trivial confirmation for our proposal.

\section{Lie algebraic structure for theories with unstable particles}

There exist some concrete Lagrangian formulations for integrable theories
with unstable particles, such as the aforementioned HSG-models (\ref{pert}).
Inspired by the structure of these models, we present here a slighly more
general Lie algebraic picture. We keep the discussion here abstract and
supply below concrete examples. For our formulation we need an arbitrary
simply laced Lie algebra $\tilde{\mathbf{g}}$ (possibly with a subalgebra $%
\tilde{\mathbf{h}}$) with rank $\tilde{\ell}$ together with its associated
Dynkin diagram (see for instance \cite{Hum}). To each node we attach a
simply laced Lie algebra $\mathbf{g}_{i}$ with rank $\ell _{i}$ and to each
link between the nodes $i$ and $j$ a resonance parameter $\sigma
_{ij}=\sigma _{i}-\sigma _{j}$, as depicted in the following $\tilde{\mathbf{%
\ g}}/\tilde{\mathbf{\ h}}$-coset Dynkin diagram

\unitlength=1.0cm 
\begin{picture}(14.20,1.4)(-2.0,-0.7)
\put(-1.,0.00){\circle*{0.2}}

\put(-1.25,-0.50){${\bold g}_1$}

\put(-0.7,0.00){$\ldots$}
\put(0,0.00){\circle*{0.2}}
\put(0.27,0.30){$ \sigma_{ij}$}

\put(-0.25,-0.50){${\bold g}_i$}
\put(0.00,-0.01){\line(1,0){0.9}}
\put(0.35,-0.01){\line(2,1){0.4}}
\put(0.35,-0.01){\line(2,-1){0.4}}
\put(1.00,0.00){\circle*{0.2}}

\put(1.35,0.30){$ \sigma_{jk}$}
\put(0.9,-0.50){${\bold g}_j$}

\put(1.0,-0.01){\line(1,0){0.9}}
\put(1.65,-0.01){\line(-2,1){0.4}}
\put(1.65,-0.01){\line(-2,-1){0.4}}
\put(2.0,0.00){\circle*{0.2}}

\put(1.9,-0.50){${\bold g}_k$}
\put(2.2,0.00){$\ldots$}

\put(3.0,0.00){\circle*{0.2}}

\put(2.9,-0.50){${\bold g}_{\tilde \ell }$}

\put(3.5,-.6){\line(1,1){1.0}}

\put(4.5,0.00){$\ldots$}
\put(5.2,0.00){\circle*{0.2}}

\put(5.55,0.30){$ \sigma_{lm}$}

\put(4.95,-0.50){${\bold g}_l$}
\put(5.20,-0.01){\line(1,0){0.9}}
\put(5.55,-0.01){\line(2,1){0.4}}
\put(5.55,-0.01){\line(2,-1){0.4}}
\put(6.20,0.00){\circle*{0.2}}
\put(6.45,0.30){$ \sigma_{mn}$}

\put(6.1,-0.50){${\bold g}_m$}

\put(6.2,-0.01){\line(1,0){0.9}}
\put(6.85,-0.01){\line(-2,1){0.4}}
\put(6.85,-0.01){\line(-2,-1){0.4}}
\put(7.2,0.00){\circle*{0.2}}
\put(7.1,-0.50){${\bold g}_n$}
\put(7.4,0.00){$\ldots$}

\end{picture}

\noindent Besides the usual rules for Dynkin diagrams, we adopt here the
convention that we add an arrow to the link, which manifests the parity
breaking and allows to identify the signs of the resonance parameters. An
arrow pointing from the node $i$ to $j$ simply indicates that $\sigma
_{ij}>0 $. Since we are dealing exclusively with simply laced Lie algebras,
this should not lead to confusion. To each simple root of the algebras $%
\mathbf{g}_{i}$, we associate now a stable particle and to each positive
non-simple root of $\tilde{\mathbf{g}}$ an unstable particle, such that 
\begin{equation}
\text{\# of stable particles}=\sum\limits_{i=1}^{\tilde{\ell}}\ell
_{i},\qquad \text{\# of unstable particles}=\frac{\tilde{\ell}\,(\tilde{h}-2)%
}{2}.  \label{nr1}
\end{equation}%
From the discussion above, we expect that the $\sigma $'s will be associated
to unstable particles, but we note that the 
\begin{equation}
\text{\# of resonance parameters}=\frac{\tilde{\ell}(\tilde{\ell}-1)}{2}
\label{nr2}
\end{equation}%
only agrees with the amount of unstable particles for $\tilde{h}=\tilde{\ell}%
+1$, e.g. for $\tilde{\mathbf{g}}=SU(\tilde{\ell}+1)$. Since the resonance
parameters govern the mass of the unstable particles, this discrepancy is
interpreted as an unavoidable mass degeneracy.

Concrete examples for this formulations are the $\tilde{\mathbf{g}}_{k}$%
-homogeneous sine-Gordon models \cite{Park:1994bx,Fernandez-Pousa:1997hi},
for which one can choose $\tilde{\mathbf{g}}$\textbf{\ }to be simply laced
and $\mathbf{g}_{1}=\ldots =\mathbf{g}_{\tilde{\ell}}=SU(k)$. This is
generalized \cite{Korff:2000zu} when taking instead $\tilde{\mathbf{g}}$%
\textbf{\ }to be non-simply laced and $\mathbf{g}_{i}=SU(2k/\alpha _{i}^{2})$%
, with $\alpha _{i}$ being the simple roots of $\tilde{\mathbf{g}}$. The
choice $\mathbf{g}_{1}=\ldots =\mathbf{g}_{\tilde{\ell}}=\mathbf{g}$ with $%
\mathbf{g}$ being any arbitrary simply laced Lie algebra gives the $\mathbf{%
g|\tilde{g}}$-theories \cite{Fring:2000ng}. An example for a theory
associated to a coset is the roaming sinh-Gordon model \cite{Stair}, which
can be thought of as $\tilde{\mathbf{\ g}}/\tilde{\mathbf{\ h}} \equiv
\lim_{k\rightarrow \infty }SU(k+1)/SU(k)$ with $\mathbf{g}_{1}=\ldots =%
\mathbf{g}_{\tilde{\ell}}=SU(2)$. It is clear that the examples presented
here do not exhaust yet all possible combinations and the structure
mentioned above allows for more combinations of algebras, which are not yet
explored. One is also not limited to Dynkin diagrams and may consider more
general graphs which have multiple links, i.e.~resonance parameters, between
various nodes. Examples for such theories were proposed and studied in \cite%
{CF88}.

\subsection{Decoupling Rule}

Of special interest is to investigate the behaviour of previously defined
systems when certain resonance parameters $\sigma $ become very large or
tend to infinity. The physical motivation for that is to describe a
renormalization group (RG) flow, which we shall discuss in more detail
below. Here we present first the mathematical set up.

\smallskip \indent {\rm{Decoupling rule}:} \emph{Call the overall Dynkin
diagram }$\mathcal{C}$\emph{\ and denote the associated Lie group and Lie
algebra by }$\tilde{G}_{\mathcal{C}}$\emph{\ and \textbf{\~{g}}}$_{\mathcal{C%
}}$\emph{, respectively. Let }$\sigma _{ij}$\emph{\ be some resonance
parameter related to the link between the nodes }$i$\emph{\ and }$j$\emph{.
To each node }$i$\emph{\ attach a simply laced Lie algebra }$\mathbf{g}_{i}.$%
\emph{\ Produce a reduced diagram }$\mathcal{C}_{ji}$\emph{\ containing the
node }$j$\emph{\ by cutting the link adjacent to it in the direction }$i$%
\emph{. Likewise produce a reduced diagram }$\mathcal{C}_{ij}$\emph{\
containing the node }$i$\emph{\ by cutting the link adjacent to it in the
direction }$j$\emph{. Then the }$\tilde{G}_{\mathcal{C}}$\emph{-theory
decouples according to the rule} 
\begin{equation}
\lim_{\sigma _{ij}\rightarrow \infty }\tilde{G}_{\mathcal{C}}=\tilde{G}_{(%
\mathcal{C}-\mathcal{C}_{ij})}\otimes \tilde{G}_{(\mathcal{C}-\mathcal{C}%
_{ji})}/\tilde{G}_{(\mathcal{C}-\mathcal{C}_{ij}-\mathcal{C}_{ji})}\,.
\label{drule}
\end{equation}
We depict this rule also graphically in terms of Dynkin diagrams:

\unitlength=1.0cm 
\begin{picture}(14.20,1.4)(0.0,-0.7)
\put(-0.7,0.00){$\ldots$}
\put(0.00,0.00){\circle*{0.2}}
\put(-0.10,-0.50){$ {\bold g}_i$}
\put(0.00,-0.01){\line(1,0){0.9}}
\put(1.00,0.00){\circle{0.2}}
\put(1.80,0.30){${\mathcal  C}$}
\put(1.10,-0.01){\line(1,0){0.5}}
\put(1.6,0.00){$\ldots$}
\put(2.2,-0.01){\line(1,0){0.4}}
\put(2.70,0.00){\circle{0.2}}
\put(2.80,-0.01){\line(1,0){1.0}}
\put(3.70,0.00){\circle*{0.2}}
\put(3.60,-0.50){${\bold g}_{j}$}
\put(3.90,-0.01){$\ldots$ }
\put(4.70,-0.1){$\Rightarrow$}
\put(4.,0.30){ \small $ \sigma_{ij} \rightarrow \infty  $}

\end{picture}

\unitlength=1.0cm 
\begin{picture}(14.20,1.4)(6.,-0.7)

\put(5.3,0.00){$\ldots$}
\put(6.00,0.00){\circle*{0.2}}
\put(5.9,-0.50){${\bold g}_i$}
\put(6.00,-0.01){\line(1,0){0.9}}
\put(7.00,0.00){\circle{0.2}}
\put(6.9,0.30){${\mathcal  C}-{\mathcal C}_{ji}$}
\put(7.10,-0.01){\line(1,0){0.5}}
\put(7.6,0.00){$\ldots$}
\put(8.2,-0.01){\line(1,0){0.4}}
\put(8.70,0.00){\circle{0.2}}

\put(9.20,-0.1){$\otimes$}

\put(10.00,0.00){\circle{0.2}}
\put(10.80,0.30){${\mathcal  C} -{\mathcal C}_{ij }$}
\put(10.10,-0.01){\line(1,0){0.5}}
\put(10.6,0.00){$\ldots$}
\put(11.2,-0.01){\line(1,0){0.4}}
\put(11.70,0.00){\circle{0.2}}
\put(11.80,-0.01){\line(1,0){1.0}}
\put(12.70,0.00){\circle*{0.2}}
\put(12.60,-0.50){${\bold g}_{j}$}
\put(12.90,-0.01){$\ldots$ }

\put(13.4,-.6){\line(1,1){1.0}}

\put(15.00,0.00){\circle{0.2}}
\put(14.90,0.30){${\mathcal  C} -{\mathcal C}_{ij} -{\mathcal C}_{ji} $}
\put(15.10,-0.01){\line(1,0){0.5}}
\put(15.6,0.00){$\ldots$}
\put(16.2,-0.01){\line(1,0){0.4}}
\put(16.70,0.00){\circle{0.2}}

\end{picture}

\noindent According to the GKO-coset construction \cite{GKO}, this means
that the Virasoro central charge flows as 
\begin{equation}
c_{\tilde{\mathbf{g}}_{\mathcal{C}}}\rightarrow c_{\tilde{\mathbf{g}}_{%
\mathcal{C-C}_{ij}}}+c_{\tilde{\mathbf{g}}_{\mathcal{C-C}_{ji}}}-c_{\mathbf{%
\tilde{g}}_{\mathcal{C}-\mathcal{C}_{ij}-\mathcal{C}_{ji}}}\,\,.  \label{dc}
\end{equation}%
The rule may be applied consecutively to each disconnected subgraph produced
according to the decoupling rule (\ref{drule}). Note that this rule
describes a decoupling and not a fusing, as it only predicts the flow in one
direction and the limit is not reversible. From a physical point of view
this is natural as also the RG flow is also irreversible. The rule (\ref%
{drule}) generalizes a rule proposed in \cite{CF8}, which was based on the
assumption that unstable particles are associated exclusively to positive
roots of height two.

More familiar in the mathematical literature is a decoupling rule found by
Dynkin \cite{Dynkin} for the construction of semi-simple\footnote{%
The subalgebras constructed in this way are not necessarily maximal and
regular. A guarantee for obtaining those, except in six special cases, is
only given when one manipulates adequately the extended Dynkin diagram.}
subalgebras $\tilde{\mathbf{h}}$ from a given algebra $\tilde{\mathbf{g}}$.
For the more general diagrams which can be related to the $\tilde{\mathbf{g}}%
_{k}$-HSG models the generalized rule can be found in \cite{Kuniba}. These
rules are all based on removing some of the nodes rather than links. For our
physical situation at hand this corresponds to sending the masses of all
stable particles which are associated to the algebra of a particular node to
infinity. As in the decoupling rule (\ref{drule}) the number of stable
particles remains preserved, it is evident that the two rules are
inequivalent. Letting for instance the mass scale in $\mathbf{g}_{j}$ go to
infinity, the generalized (in the sense that $\mathbf{g}_{j}$ can be
different from $A_{\ell }$) rule of Kuniba is simply depicted as

\unitlength=1.0cm 
\begin{picture}(14.20,1.4)(0.0,-0.7)
\put(-0.7,0.00){$\ldots$}
\put(0.00,0.00){\circle*{0.2}}
\put(-0.10,-0.50){$ {\bold g}_i$}
\put(0.00,-0.01){\line(1,0){0.9}}
\put(1.00,0.00){\circle*{0.2}}

\put(0.90,-.50){$ {\bold g}_j$}

\put(0.9,0.30){${\mathcal  C}$}
\put(1.10,-0.01){\line(1,0){0.9}}
\put(2.00,0.00){\circle*{0.2}}
\put(1.90,-0.50){$ {\bold g}_k$}
\put(2.20,-0.01){$\ldots$ }

\put(3.80,-0.1){$\Rightarrow$}
\put(3.1,0.30){ \small $ m_j \rightarrow \infty  $}

\put(5.3,0.00){$\ldots$}
\put(6.00,0.00){\circle*{0.2}}
\put(5.9,-0.50){${\bold g}_i$}
\put(5.5,0.30){${\mathcal  C}-{\mathcal C}_{ji}$}

\put(7.50,-0.1){$\otimes$}

\put(8.50,0.30){${\mathcal  C} -{\mathcal C}_{jk} $}

\put(9.0,0.00){\circle*{0.2}}
\put(8.90,-0.50){${\bold g}_{k}$}
\put(9.20,-0.01){$\ldots$ }
\end{picture}

\noindent Clearly this can not be produced with (\ref{drule}).

\subsection{A simple example: The SU(4)$_{2}$-HSG model}

We illustrate the working of the rule (\ref{drule}) with a simple example.
We take $\tilde{\mathbf{g}}$ to be $SU(4)$, attach to each node simply an $%
SU(2)$ algebra and to the links the resonance parameters $\sigma
_{12},\sigma _{13},\sigma _{23}$. This corresponds to the $SU(4)_{2}$-HSG
model. For the ordering $\sigma _{13}>\sigma _{12}>\sigma _{23}$ the rule (%
\ref{drule}) predicts the following flow

\unitlength=1.0cm 
\begin{picture}(15.20,4.0)(-0.9,-2.5)
\put(0,1.00){\circle*{0.2}}
\put(-0.25,.50){$\alpha_1$}
\put(0.00,0.99){\line(1,0){0.9}}
\put(1.00,1.00){\circle*{0.2}}
\put(0.9,0.50){$\alpha_2$}
\put(1.00,0.99){\line(1,0){0.9}}
\put(2.00,1.00){\circle*{0.2}}
\put(1.9,0.50){$\alpha_3$}
\put(6.3,0.85){\footnotesize ${\bf \tilde{g}} = SU(4)_2$}
\put(9.3,0.8){$c = 2$}
\put(-1.5,0.00){$ \rightarrow \sigma_{13}$}
\put(0,0.00){\circle*{0.2}}
\put(-0.25,-0.50){$\alpha_1$}
\put(0.00,-0.01){\line(1,0){0.9}}
\put(1.00,0.00){\circle*{0.2}}
\put(0.9,-0.50){$\alpha_2$}
\put(1.50,-0.1){$\otimes$}
\put(2.1,0.0){\circle*{0.2}}
\put(1.95,-0.50){$\alpha_2$}
\put(2.10,-0.01){\line(1,0){0.9}}
\put(3.10,0.00){\circle*{0.2}}
\put(3.,-0.50){$\alpha_3$}
\put(3.6,-.6){\line(1,1){1.0}}
\put(4.90,0.00){\circle*{0.2}}
\put(4.8,-.50){$\alpha_2$}
\put(6.3,-0.15){\footnotesize ${\bf \tilde{g}} = SU(3)^{\otimes 2}_2/SU(2)_2$}
\put(9.3,-0.2){$c = 1.9$}
\put(-1.5,-1.00){$ \rightarrow  \sigma_{12}$}
\put(.00,-1.00){\circle*{0.2}}
\put(-0.25,-1.50){$\alpha_1$}
\put(.50,-1.1){$\otimes$}
\put(1.1,-1.0){\circle*{0.2}}
\put(0.95,-1.50){$\alpha_2$}
\put(1.10,-1.01){\line(1,0){0.9}}
\put(2.10,-1.00){\circle*{0.2}}
\put(2.,-1.50){$\alpha_3$}
\put(6.3,-1.15){\footnotesize ${\bf \tilde{g}} = SU(3)_2 \otimes SU(2)_2 $} 
\put(9.3,-1.2){$c = 1.7$}
\put(-1.5,-2.00){$ \rightarrow \sigma_{23}$}
\put(0,-2.00){\circle*{0.2}}
\put(-0.25,-2.50){$\alpha_1$}
\put(.40,-2.1){$\otimes$}
\put(1.00,-2.00){\circle*{0.2}}
\put(0.9,-2.50){$\alpha_2$}
\put(1.50,-2.1){$\otimes$}
\put(2.1,-2.0){\circle*{0.2}}
\put(1.95,-2.50){$\alpha_3$}
\put(6.3,-2.15){\footnotesize ${\bf \tilde{g}} = SU(2)^{\otimes 3}_2 $}
\put(9.3,-2.2){$c = 1.5$}
\end{picture}

\noindent The central charges are obtained from (\ref{cdel}) using (\ref{dc}%
). Chosing instead the ordering $\sigma _{23}>\sigma _{13}>\sigma _{12}$, we
compute

\begin{picture}(15.20,3.8)(-0.9,-2.5)

\put(0,1.00){\circle*{0.2}}
\put(-0.25,.50){$\alpha_1$}
\put(0.00,0.99){\line(1,0){0.9}}
\put(1.00,1.00){\circle*{0.2}}
\put(0.9,0.50){$\alpha_2$}
\put(1.00,0.99){\line(1,0){0.9}}
\put(2.00,1.00){\circle*{0.2}}
\put(1.9,0.50){$\alpha_3$}
\put(6.3,0.85){\footnotesize ${\bf \tilde{g}} = SU(4)_2$}
\put(9.3,0.8){$c = 2$}
\put(-1.5,0.00){$\rightarrow \sigma_{23}$}
\put(0,0.00){\circle*{0.2}}
\put(-0.25,-0.50){$\alpha_1$}
\put(0.00,-0.01){\line(1,0){0.9}}
\put(1.00,0.00){\circle*{0.2}}
\put(0.9,-0.50){$\alpha_2$}
\put(1.50,-0.1){$\otimes$}
\put(2.1,0.0){\circle*{0.2}}
\put(1.95,-0.50){$\alpha_3$}
\put(6.3,-0.15){\footnotesize ${\bf \tilde{g}} = SU(3)_2 \otimes SU(2)_2$}
\put(9.3,-0.2){$c = 1.7$}
\put(-1.5,0.00){$\rightarrow \sigma_{23}$}
\put(0,0.00){\circle*{0.2}}
\put(-0.25,-0.50){$\alpha_1$}
\put(0.00,-0.01){\line(1,0){0.9}}
\put(1.00,0.00){\circle*{0.2}}
\put(0.9,-0.50){$\alpha_2$}
\put(1.50,-0.1){$\otimes$}
\put(2.1,0.0){\circle*{0.2}}
\put(1.95,-0.50){$\alpha_3$}
\put(6.3,-0.15){\footnotesize ${\bf \tilde{g}} = SU(3)_2 \otimes SU(2)_2$}
\put(9.3,-0.2){$c = 1.7$}
\put(-1.5,-1.00){$ \rightarrow \sigma_{13}$}
\put(.00,-1.025){is already decoupled}
\put(-1.5,-2.00){$ \rightarrow \sigma_{12}$}
\put(0,-2.00){\circle*{0.2}}
\put(-0.25,-2.50){$\alpha_1$}
\put(.40,-2.1){$\otimes$}
\put(1.00,-2.00){\circle*{0.2}}
\put(0.9,-2.50){$\alpha_2$}
\put(1.50,-2.1){$\otimes$}
\put(2.1,-2.0){\circle*{0.2}}
\put(1.95,-2.50){$\alpha_3$}
\put(6.3,-2.15){\footnotesize ${\bf \tilde{g}} = SU(2)^{\otimes 3}_2 $}
\put(9.3,-2.2){$c = 1.5$}

\end{picture}

\noindent It is important to note the non-commutative nature of the limiting
procedures. For more complicated algebras it is essential to keep track of
the labels on the nodes, since only in this way one can decide whether they
cancel against the subgroup diagrams or not.

\subsection{A non-trivial example: The (E$_{6}$)$_{2}$-HSG model}

As by now we do not have a rigorous proof of the decoupling rule (\ref{drule}%
), we take the support for its validity from the working of various
examples. We will check below the analytic predictions of the rule against
some alternative method. As the previous example was a simple pedagogical
one, we will consider next a non-trivial one leading to an intricate
prediction for the RG-flow. The confirmative double check below can hardy be
accidental and we take that as very strong support for the validity of (\ref%
{drule}).

We consider now the (E$_{6}$)$_{2}$-HSG model. In this case we have $\tilde{%
\ell}=6,\tilde{h}=12$ such that we have $6$ stable particles, $30$ unstable
particles and $15$ resonance parameters. From the $5!$ possible orderings
for the resonance parameters we present here only two concrete ones, which
will predict different types of flows and mass degeneracies. Note that this
degeneracy is not the unavoidable one resulting from the difference between
the number of resonance parameters and non-simple positive roots that is $%
30-15$, as discussed for (\ref{nr1}) and (\ref{nr2}). The degeneracies
discussed here are a consequence of the particular choices of the resonance
parameters. The conventions for the labeling of our particles are indicated
in the following Dynkin diagram:

\unitlength=1.cm 
\begin{picture}(14.20,2.2)(-1.5,0.2)
\put(0,1.00){\circle*{0.2}}
\put(-0.25,.50){$\alpha_1$}
\put(0.00,.99){\line(1,0){0.9}}
\put(1.00,1.00){\circle*{0.2}}
\put(0.9,0.50){$\alpha_3$}
\put(1.00,0.99){\line(1,0){0.9}}
\put(2.00,1.00){\circle*{0.2}}
\put(1.9,0.50){$\alpha_4$}
\put(2.02,0.99){\line(0,1){0.9}}
\put(2.00,1.99){\circle*{0.2}}
\put(2.2,1.8){$\alpha_2$}
\put(2.00,0.99){\line(1,0){0.9}}
\put(3.00,1.00){\circle*{0.2}}
\put(2.9,0.50){$\alpha_5$}
\put(3.00,0.99){\line(1,0){0.9}}
\put(4.00,1.00){\circle*{0.2}}
\put(3.9,0.50){$\alpha_6$}
\end{picture}

\noindent We choose first the ordering and values for resonance parameters
as 
\begin{equation}
\sigma _{13}=100>\sigma _{34}=80>\sigma _{45}=60>\sigma _{56}=40>\sigma
_{24}=20\,.
\end{equation}%
According to the decoupling rule (\ref{drule}), we predict therefore the
flow:

\begin{equation*}
\begin{array}{lll}
& E_{6} & \frac{36}{7}\sim 5.\,\allowbreak 14 \\ 
\rightarrow \sigma _{16}=280 & SO(10)^{\otimes 2}/SO(8) & 5 \\ 
\rightarrow \sigma _{15}=240 & SO(10)\otimes SU(5)/SU(4) & \frac{34}{7}\sim
4.\,\allowbreak 86 \\ 
\rightarrow \sigma _{14},\sigma _{36}=180\quad & SO(8)\otimes SU(5)\otimes
SU(3)/SU(4)\otimes SU(2) & \frac{319}{70}\sim 4.\,\allowbreak 56 \\ 
\rightarrow \sigma _{12}=160 & \text{is already decoupled} &  \\ 
\rightarrow \sigma _{35}=140 & SU(5)\otimes SU(4)\otimes SU(3)/SU(3)\otimes
SU(2) & \frac{61}{14}\sim 4.\,\allowbreak 36 \\ 
\rightarrow \sigma _{26}=120 & SU(4)^{\otimes 3}\otimes SU(3)/SU(3)^{\otimes
2}\otimes SU(2) & 4.3 \\ 
\rightarrow \sigma _{13},\sigma _{46}=100 & SU(4)^{\otimes 2}\otimes
SU(3)\otimes SU(2)/SU(3)\otimes SU(2)\quad & 4 \\ 
\rightarrow \sigma _{25},\sigma _{34}=80 & SU(3)^{\otimes 3}\otimes
SU(2)^{\otimes 2}/SU(2)^{\otimes 2} & 3.6 \\ 
\rightarrow \sigma _{32},\sigma _{45}=60 & SU(3)^{\otimes 2}\otimes
SU(2)^{\otimes 2} & 3.4 \\ 
\rightarrow \sigma _{56}=40 & SU(3)\otimes SU(2)^{\otimes 4} & 3.2 \\ 
\rightarrow \sigma _{24}=20 & SU(2)^{\otimes 6} & 3%
\end{array}%
\end{equation*}

\noindent\ Note that eight particles are pairwise degenerate and we
therefore expect to find $15-8/2=11$ plateaux in the flow. The first step
which corresponds to one of these degeneracies occurs for instance at $%
\sigma _{14}=\sigma _{36}$ and we have to apply the decoupling rule twice at
this point before we get a new fixed point theory.

Next we arrange the couplings as 
\begin{equation}
\sigma _{45}=100>\sigma _{34}=80>\sigma _{13}=60>\sigma _{56}=40>\sigma
_{24}=20\,.
\end{equation}

\noindent and compute from (\ref{drule}) the flow 
\begin{equation*}
\begin{array}{lll}
& E_{6} & \frac{36}{7}\sim 5.\,\allowbreak 14 \\ 
\rightarrow \sigma _{16}=280 & SO(10)^{\otimes 2}/SO(8) & 5 \\ 
\rightarrow \sigma _{15}=240 & SO(10)\otimes SU(5)/SU(4) & \frac{34}{7}\sim
4.\,\allowbreak 86 \\ 
\rightarrow \sigma _{36}=220 & SU(5)^{\otimes 2}\otimes SO(8)/SU(4)^{\otimes
2} & \frac{33}{7}\sim 4.\,\allowbreak 71 \\ 
\rightarrow \sigma _{35}=180 & SU(5)^{\otimes 2}/SU(3) & \frac{158}{35}\sim
4.\,\allowbreak 51 \\ 
\rightarrow \sigma _{26}=160 & SU(4)^{\otimes 2}\otimes SU(5)/SU(3)^{\otimes
2} & \frac{156}{35}\sim 4.\,\allowbreak 46 \\ 
\rightarrow \sigma _{14}=\sigma _{46}=140\quad & SU(4)^{\otimes 2}\otimes
SU(3)^{\otimes 2}/SU(2)^{\otimes 2}\otimes SU(3)\quad & 4.2 \\ 
\rightarrow \sigma _{12}=\sigma _{25}=120 & SU(4)\otimes SU(3)^{\otimes
3}/SU(2)^{\otimes 3} & 4.1 \\ 
\rightarrow \sigma _{45}=100 & SU(4)\otimes SU(3)^{\otimes 2}/SU(2) & 3.9 \\ 
\rightarrow \sigma _{34}=80 & SU(3)^{\otimes 3} & 3.6 \\ 
\rightarrow \sigma _{13}=\sigma _{32}=60 & SU(3)^{\otimes 2}\otimes
SU(2)^{\otimes 2} & 3.4 \\ 
\rightarrow \sigma _{56}=40 & SU(3)\otimes SU(2)^{\otimes 4} & 3.2 \\ 
\rightarrow \sigma _{24}=20 & SU(2)^{\otimes 6} & 3%
\end{array}%
\end{equation*}%
In this case we have only six particles pairwise degenerate and we expect to
find $15-6/2=12$ plateaux. In the next section we find that the predictions
made here are confirmed even for this involved case.

\section{How to detect unstable particles?}

In section 2 we described several arguments which predict the spectrum of
unstable particles and now we will present some methods which allow to test
these predictions. In particular with regard to the boostrap proposal this
will be important, as it is not yet rigorously supported. Computing
renormalization group (RG) flows will allow to detect the unstable
particles. Roughly speaking, the central idea of an RG analysis is to probe
different energy scales of a theory. We can flow from an energy regime so
large that the unstable particle can energetically not exist to one in which
it is formed. As a consequence, the particle content of the theory will be
altered, which is visible in form of a typical staircase pattern of the RG
scaling function.

There are various ways to compute such scaling functions, such as the
evaluation of the c-theorem \cite{cZam} or an analysis by means of the
thermodynamic Bethe ansatz (TBA) \cite{Zamolodchikov:1990cf}. In the first
case we have to evaluate the expression%
\begin{equation}
c(r_{0})=\frac{3}{2}\int\limits_{r_{0}}^{\infty }dr\,r^{3}\,\,\left\langle
\Theta (r)\Theta (0)\right\rangle \,\,.  \label{cth}
\end{equation}%
The main difficulty in this approach is to evaluate the two-point
correlation function $\,\left\langle \Theta (r)\Theta (0)\right\rangle $ for
the trace of the energy-momentum tensor $\Theta $ depending on the radial
distance $r$. Most effectively, one can do this by expanding it in terms of
\ form factors, for a general recent introduction see e.g. \cite{SG} and
references therein. It is well known that for many, even quite non-trivial,
theories such form factor expansions converge extremely fast, see \cite{Ren}
for the compution of (\ref{cth}) for the SU(3)$_{2}$-HSG model.

Here we will concentrate more on the TBA, which is simpler to handle in most
cases. As a prerequisite, one assumes to know all scattering matrices $%
S_{ij}(\theta )$ for the \textbf{stable} particles of the type $i$,$j$ with
masses $m_{i},m_{j}$. Besides this dynamical interaction one also makes an
assumption on the statistical interaction between the particles, which are
choosen here to be of fermionic type. The TBA consists now of compactifying
the space of this 1+1 dimensional relativistic model into a circle of finite
circumference $R$, such that all energies become discrete and functions of $%
R $. The function similar to (\ref{cth}), which scales now these energies
takes on the form 
\begin{equation}
c_{\text{eff}}(r)=\frac{3\,r}{\pi ^{2}}\sum_{i}m_{i}\int\limits_{-\infty
}^{\infty }d\theta \,\cosh \theta \,\ln (1+e^{-\varepsilon _{i}(\theta
,r)})\,.  \label{scale}
\end{equation}%
One identifies the circumference $R$ with the inverse temperature $T$ and
introduces the scaling parameter $r=m/T$, \ with $m$ being an overall mass
scale. The $\varepsilon _{i}(\theta ,r)$ are the so-called the
pseudo-energies which can be obtained as solutions of the thermodynamic
Bethe ansatz equations 
\begin{equation}
rm_{i}\cosh \theta =\varepsilon _{i}(\theta ,r)+\sum\limits_{j}[\varphi
_{ij}\ast \ln (1+e^{-\varepsilon _{j}})](\theta ,r)\,.  \label{TBA}
\end{equation}%
Here the $\ast $ denotes the convolution of two functions $\left( f\ast
g\right) (\theta )$ $:=1/(2\pi )\int d\theta ^{\prime }f(\theta -\theta
^{\prime })g(\theta ^{\prime })$ and the $S$ (for the stable particles
only!) enter via their logarithmic derivatives $\varphi _{ij}(\theta
)=-id\ln S_{ij}(\theta )/d\theta $. The main difficulty in this approach is
to solve (\ref{TBA}), which are coupled non-linear integral equations and
therefore not solvable in a systematic analytical way.

Now it is clear, that the two functions (\ref{cth}) and (\ref{scale}) can
not be the same, but nevertheless they contain the same qualitative
information. The functions will flow through various fixed points, at which
the theory become effectively conformal field theories and the
normalizations are choosen in such a way that the values of both functions
coincide with the corresponding Virasoro central charges. When the theory is
not unitary, (\ref{scale}) has to be corrected by an additive term to
achieve this. Computing then a flow from the infrared to the ultraviolet,
one passes now various CFT plateaux, where the changes are associated to the
formation of unstable particles with mass (\ref{m}). The challenge is of
course to predict the positions, that is, the height and the on-set of the
plateaux, as a function of the scaling parameter. The on-set is related to
the energy scale of the unstable particles and thus simply determined by the
formula (\ref{m}). To predict the height is less trivial and the proposal
made in \cite{C30F} is that the decoupling rule (\ref{drule}) achieves this.
It is important to note here that $\sigma \rightarrow \infty $ in (\ref%
{drule}), means in the RG context $\sigma \gg $ all other resonance
parameters. In the follwing picture we present the numerical computation for
the (E$_{6}$)$_{2}$-HSG model, which precisely confirms our analytical
predictions made by the decoupling rule in section 3.3

\epsfig{file=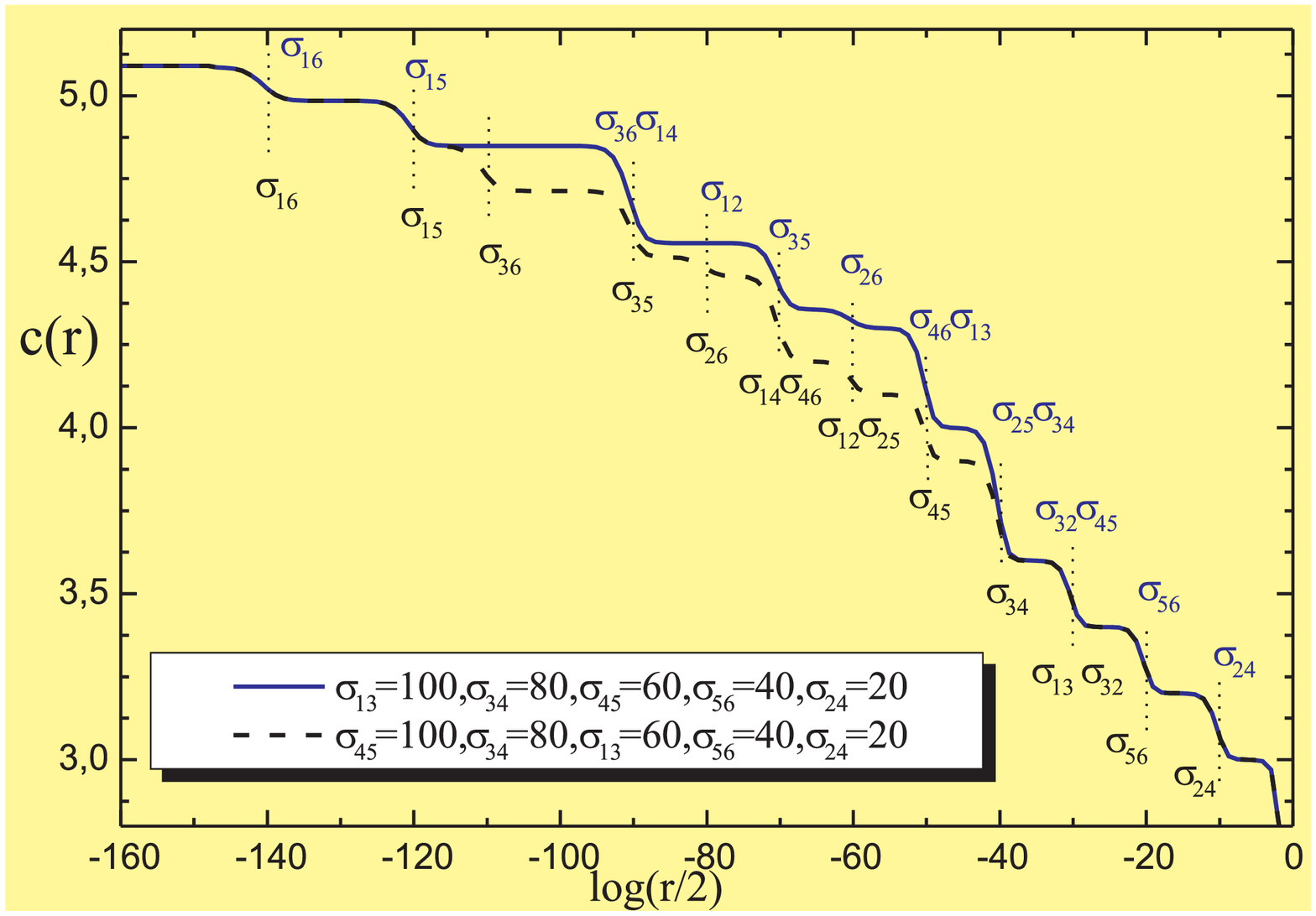,width=88mm}

Having confirmed the predictions of our decoupling rule with a TBA-analysis,
let us now discuss how the results of this analysis are compatible with our
bootstrap proposal with a simple example: We consider the processes (\ref{27}%
), (\ref{28}). In order to be able to interpret the BW-formula for the
production of the particle (123) from (12) + 3 or (23) + 1 one has to
\textquotedblleft make\textquotedblright\ (12) and (23) stable, which is
achieved when $\sigma _{12}$ or $\sigma _{23}$ is zero. One has to do that
as otherwise the BW can not be applied, it only makes sense for stable
particles. The first not obvious result here is that the resulting mass for
(123) turns out to be the same from both cases (\ref{28}) (and in all other
examples!!!). Looking at the outcome of the TBA calculation (see \cite{C30F}
for the numerics on this case) one finds precisely the value (\ref{m13})
reproduced by the TBA at the onset $\ln (r/2)\sim -\sigma _{13}/2=-(\sigma
_{12}+\sigma _{23})/2$. Now apparently in the TBA analysis $\sigma _{12}$ or 
$\sigma _{23}$ are not zero, which seems to contradict the previous
assumptions in the bootstrap analysis. To understand this, one should keep
in mind the meaning of the steps in the TBA. The formation of the particle
(123) takes place when its mass becomes greater than the energy scale of the
RG-flow, i.e. when $m\exp (\sigma _{13}/2)>2m/r$. Let us chose for instance $%
\sigma _{12}=30$, $\sigma _{23}=60$, then $\exp (\sigma _{13}/2)\sim \exp
(45)\sim 3.49\times 10^{19}$. To resolve the apparent contradiction, it is
now important to note that the other unstable particles are formed several
orders of magnitude below at $\exp (30)\sim 1.06\times 10^{13}$ and $\exp
(15)\sim 3.72\times 10^{6}$. This means in comparison to the formation
energy scale of particle (123) the parameters $\sigma _{12}$, $\sigma _{23}$%
\ can be regarded as approximately zero, which is in agreement with the
assumption in the bootstrap analysis!

This is just the same picture as put forward in the decoupling rule: In the
formulation we say $\sigma _{13}\rightarrow \infty $, but inside the TBA
analysis this is a milder statement and just means $\sigma _{13}\gg \sigma
_{12},\sigma _{23}$. Further quite non-obvious confirmation comes from the
results when choosing the parameters differently, i.e. in the example
discussed here $\sigma _{12}\rightarrow -\sigma _{12}$. The two pictures
completely coincide. With regard to previous studies, it is very important
to note that the occurrence of the step at $\exp (\sigma _{13}/2)\sim 2/r$
had no explanation at all before. Only the onsets at $\exp (\sigma
_{23}/2)\sim 2/r$ and $\exp (\sigma _{12}/2)\sim 2/r$ could be explained as
they correspond to the formation of unstable particles from two stable ones.
The additional step (for other algebras there are far more) was a mystery
pointed out first in \cite{Dorey:2002sc}. In \cite{C30F} we provided for the
first time an explanation for this feature: We predict its height and
on-set, thus explaining also why it is absent when the resonance parameters
are chosen differently. For all other examples studied (not even all have
been presented in this proceeding, see \cite{C30F} for more) this picture is
completely consistent.

\section{Theories with an infinite amount of unstable particles}

We address now the question of how to enlarge a given finite particle
spectrum of a theory to an infinite one. In general the bootstrap (\ref{boot}%
), which is the central construction principle for the $S$-matrix, is
assumed to close after a finite number of steps, which means it involves a
finite number of particles. However, from a physical as well as from a
mathematical point of view, it appears to be natural to extend the
construction in such a way that it would involve an infinite number of
particles. The physical motivation for this are string theories, which admit
an infinite particle spectrum. Mathematically the infinite bootstrap would
be an analogy to infinite dimensional groups, in the sense that two entries
of the S-matrix are combined into a third, which is again a member of the
same infinite set. It appears to us that it is impossible to construct an
infinite bootstrap system involving asymptotic states, although we do not
know a rigorous proof of such a no-go theorem. Instead, we will demonstrate
that it is possible to introduce an infinite number of unstable particles
into the spectrum.

\subsection{q-deformed gamma functions and Jacobian elliptic functions}

In general, the $S$-matrix amplitudes consist of (in)finite products of
hyperbolic or/and gamma functions. Here we will argue, that to enlarge the
spectrum to an infinite number one should replace these functions by
q-deformed quantities or elliptic functions. Let us first recall some
mathematical facts in this section. We start with some properties of
q-deformed quantities, which have turned out to be very useful objects as
they allow for instance to carry out elegantly (semi)-classical limits when
the deformation parameter is associated to Planck's constant. Here we define
a deformation parameter $q$ and its Jacobian imaginary transformed version,
i.e.~$\tau \rightarrow -1/\tau $, as 
\begin{equation}
q=\exp (i\pi \tau ),\qquad \hat{q}=\exp (-i\pi /\tau ),\qquad \tau
=iK_{1-\ell }/K_{\ell }~~.  \label{q}
\end{equation}%
We introduced here the quarter periods $K_{\ell }$ of the Jacobian elliptic
functions depending on the parameter $\ell \in \lbrack 0,1]$, defined in the
usual way through the complete elliptic integrals $K_{\ell }=\int_{0}^{\pi
/2}(1-\ell \sin ^{2}\theta )^{-1/2}d\theta \,$. Then 
\begin{equation}
\lim_{\ell \rightarrow 0,\hat{q}\rightarrow 1}K_{\ell }=\lim_{\ell
\rightarrow 1,q\rightarrow 1}K_{1-\ell }=\pi /2\text{,}\qquad \text{\ }%
\lim_{\ell \rightarrow 0,\hat{q}\rightarrow 1}K_{1-\ell }=\lim_{\ell
\rightarrow 1,q\rightarrow 1}K_{\ell }\rightarrow \infty ~.  \label{kk}
\end{equation}%
It will turn out below that quantities in $\hat{q}$ will be most relevant
for our purposes and therefore we state several identities directly in $\hat{%
q}$, rather than $q$, even when they hold for generic values. The most basic
q-deformed objects one defines are q-deformed integers (numbers), for which
we take the convention 
\begin{equation}
\lbrack n]_{\hat{q}}:=\frac{\hat{q}^{n}-\hat{q}^{-n}}{\hat{q}-\hat{q}^{-1}}%
.\qquad \quad
\end{equation}%
They have the obvious properties 
\begin{eqnarray}
\lim_{\ell \rightarrow 0}[n]_{\hat{q}} &=&n,  \label{limn} \\
\lim_{\ell \rightarrow 0}\frac{[n+m\tau ]_{\hat{q}}}{[n^{\prime }+m^{\prime
}\tau ]_{\hat{q}}} &=&\left\{ 
\begin{array}{c}
1~~~~~~\qquad ~~\text{for~~~~}m,m^{\prime }\neq 0 \\ 
n/n^{\prime }\qquad ~~~~\ ~\text{for~~~~}m=m^{\prime }=0%
\end{array}%
\right. ~.  \label{limnn}
\end{eqnarray}%
Next we define a q-deformed version of Euler's gamma function 
\begin{equation}
\Gamma _{\hat{q}}(x+1):=\prod\limits_{n=1}^{\infty }\frac{[1+n]_{\hat{q}%
}^{x}[n]_{\hat{q}}}{[x+n]_{\hat{q}}[n]_{\hat{q}}^{x}}~.\qquad  \label{gq}
\end{equation}%
The crucial property of the function $\Gamma _{\hat{q}}$, which coins also
its name, is 
\begin{equation}
\lim_{\ell \rightarrow 0}\Gamma _{\hat{q}}(x+1)=\lim_{\hat{q}\rightarrow
1}\Gamma _{\hat{q}}(x+1)=\prod\limits_{n=1}^{\infty }\frac{n}{n+x}\left( 
\frac{1+n}{n}\right) ^{x}=\Gamma (x+1)~.\qquad  \label{gqn}
\end{equation}%
We can relate deformations in $q$ and $\hat{q}$ through 
\begin{equation}
\frac{\hat{q}^{(x+\tau /2-1/2)^{2}}}{\hat{q}^{(y+\tau /2-1/2)^{2}}}\frac{%
\Gamma _{\hat{q}}(y)\Gamma _{\hat{q}}(1-y)}{\Gamma _{\hat{q}}(x)\Gamma _{%
\hat{q}}(1-x)}=\frac{\Gamma _{q}(-y/\tau )\Gamma _{q}(1+y/\tau )}{\Gamma
_{q}(-x/\tau )\Gamma _{q}(1+x/\tau )}~.
\end{equation}%
Frequently we have to shift the argument by integer values 
\begin{equation}
\Gamma _{\hat{q}}(x+1)=\hat{q}^{x-1}[x]_{\hat{q}}\Gamma _{\hat{q}}(x)~.
\label{11}
\end{equation}%
Relation (\ref{11}) can be obtained directly from (\ref{gq}). As a
consequence of this we also have 
\begin{eqnarray}
\Gamma _{\hat{q}}(x+m) &=&\Gamma _{\hat{q}}(x)\prod\limits_{l=0}^{m-1}\hat{q}%
^{x+l-1}[x+l]_{\hat{q}}\quad \qquad \quad \quad ~~~~m\in \mathbb{Z}
\label{33} \\
\Gamma _{\hat{q}}(x) &=&\Gamma _{\hat{q}}(x-m)\prod\limits_{l=0}^{m-1}\hat{q}%
^{x-l-2}[x-l-1]_{\hat{q}}\quad \quad m\in \mathbb{Z~}.  \label{333}
\end{eqnarray}%
Whereas (\ref{11})-(\ref{33}) hold for generic $q$, the following identities
are only valid for $\hat{q}$ 
\begin{eqnarray}
\Gamma _{\hat{q}}(1/2-\tau /2)\Gamma _{\hat{q}}(1/2+\tau /2) &=&\ell
^{1/4}\Gamma _{\hat{q}}(1/2)^{2}  \label{2} \\
\frac{\Gamma _{\hat{q}}(x+2\tau )}{\Gamma _{\hat{q}}(y+2\tau )} &=&\frac{%
\Gamma _{\hat{q}}(x)}{\Gamma _{\hat{q}}(y)}  \label{ola1} \\
\prod\limits_{i=1}^{p}\frac{\Gamma _{\hat{q}}(x_{i})\Gamma _{\hat{q}%
}(x_{i}\pm \tau /2)}{\Gamma _{\hat{q}}(y_{i})\Gamma _{\hat{q}}(y_{i}\pm \tau
/2)} &=&\prod\limits_{i=1}^{p}\frac{\Gamma _{\hat{q}^{2}}(x_{i})}{\Gamma _{%
\hat{q}^{2}}(y_{i})}\quad ~~~\text{if\quad\ }\sum\limits_{i=1}^{p}x_{i}=\sum%
\limits_{i=1}^{p}y_{i}~  \label{4p} \\
\lim_{\hat{q}\rightarrow 1}\prod\limits_{i=1}^{p}\frac{\Gamma _{\hat{q}%
}(x_{i}\pm \tau /2)}{\Gamma _{\hat{q}}(y_{i}\pm \tau /2)} &=&1\quad \qquad
\qquad ~~~~~\text{if\quad\ }\sum\limits_{i=1}^{p}x_{i}=\sum%
\limits_{i=1}^{p}y_{i}~~~~  \label{5p} \\
\lim_{\hat{q}\rightarrow 1}\frac{1}{\ell ^{1/4}}\Gamma _{\hat{q}}(\frac{x}{%
2K_{\ell }}\mp \frac{\tau }{2})\Gamma _{\hat{q}}(1-\frac{x}{2K_{\ell }}\pm 
\frac{\tau }{2}) &=&\pi ~\qquad \qquad ~\text{for \quad ~~~}x\neq 0  \notag
\label{4}
\end{eqnarray}%
Most of these properties can be checked directly by means of the defining
relation (\ref{gq}). \ The singularity structure will be important for the
physical applications. It follows from (\ref{gq}) that the $\Gamma _{\hat{q}%
} $-function\ has no zeros, but poles 
\begin{equation}
\lim_{\theta \rightarrow \theta _{\Gamma ,p}^{nm}=m\tau -n}\Gamma _{\hat{q}%
}(\theta +1)\rightarrow \infty \qquad \quad \text{for~~}~m\in \mathbb{Z}%
,n\in \mathbb{N}\mathbf{~.}  \label{pG}
\end{equation}%
Next we define 
\begin{equation}
\{x\}_{\theta }^{\sigma }\,:=\frac{\tanh (\theta -i\pi x+\sigma )/2}{\tanh
(\theta +i\pi x+\sigma )/2},\quad \{x\}_{\theta ,\ell }^{\sigma
}:=\prod_{n=-\infty }^{\infty }\{x\}_{\theta -n\log q}^{\sigma }=\frac{\func{%
sc}\theta _{-}\func{dn}\theta _{+}}{\func{sc}\theta _{+}\func{dn}\theta _{-}}
\label{gen}
\end{equation}%
with $x\in \mathbb{Q}$, $\sigma \in \mathbb{R}$ and $\theta _{\pm }=(\theta
\pm i\pi x+\sigma )iK_{\ell }/\pi $. We employed here the Jacobian elliptic
functions for which we use the common notation $\func{pq}(z)$ with p,q $\in
\{$s,c,d,n$\}$ (see e.g.~\cite{Elliptic} for standard properties). We derive
important relations between the q-deformed gamma functions and the Jacobian
elliptic $\func{sn}$-function

\begin{eqnarray}
\func{sn}(x) &=&\frac{1}{\ell ^{\frac{1}{4}}}\frac{\Gamma _{\hat{q}}(\frac{x%
}{2K_{\ell }}\mp \frac{\tau }{2})\Gamma _{\hat{q}}(1-\frac{x}{2K_{\ell }}\pm 
\frac{\tau }{2})}{\Gamma _{\hat{q}}(\frac{x}{2K_{\ell }})\Gamma _{\hat{q}}(1-%
\frac{x}{2K_{\ell }})},  \label{sng} \\
&=&\frac{q^{\frac{1}{4}-\frac{ix}{2K_{1-\ell }}}}{i\ell ^{\frac{1}{4}}}\frac{%
\Gamma _{q}(\frac{1}{2}+\frac{ix}{2K_{1-\ell }})\Gamma _{q}(\frac{1}{2}-%
\frac{ix}{2K_{1-\ell }})}{\Gamma _{q}(1-\frac{ix}{2K_{1-\ell }})\Gamma _{q}(%
\frac{ix}{2K_{1-\ell }})}~.  \label{sng2}
\end{eqnarray}%
These relations can be used to obtain some of the above mentioned
expressions. For instance, recalling that $\func{sn}(K_{\ell })=1$, we
obtain (\ref{2}). With (\ref{gq}) we recover from this the well known
identity $\func{sn}(iK_{1-\ell }/2)=i/\ell ^{1/4}$. The trigonometric limits 
\begin{eqnarray}
\lim_{\ell \rightarrow 0}\func{sn}(x) &=&\lim_{\hat{q}\rightarrow 1}\func{sn}%
(x)=\frac{\pi }{\Gamma (\frac{x}{\pi })\Gamma (1-\frac{x}{\pi })}=\sin (x)
\label{ts} \\
\lim_{\ell \rightarrow 1}\func{sn}(x) &=&\lim_{q\rightarrow 1}\func{sn}(x)=%
\frac{1}{i}\frac{\Gamma (\frac{1}{2}+\frac{ix}{\pi })\Gamma (\frac{1}{2}-%
\frac{ix}{\pi })}{\Gamma (1-\frac{ix}{\pi })\Gamma (\frac{ix}{\pi })}=\tanh
(x).
\end{eqnarray}%
can be read off directly recalling (\ref{kk}), (\ref{gqn}) and presuming
that (\ref{4}) holds. We recall the zeros and poles of the Jacobian elliptic 
$\func{sn}(\theta )$-function, which in our conventions are located at 
\begin{eqnarray}
\text{zeros}\text{:\qquad } &&\theta _{\func{sn},0}^{lm}=2lK_{\ell
}+i2mK_{1-\ell }\qquad \quad \quad \quad l,m\in \mathbb{Z}  \label{zJ} \\
\text{poles}\text{:\qquad } &&\theta _{\func{sn},p}^{lm}=2lK_{\ell
}+i(2m+1)K_{1-\ell }\qquad \qquad ~~l,m\in \mathbb{Z}\mathbf{~.}  \label{pJ}
\end{eqnarray}%
We have now assembled the main properties of the q-deformed functions which
we shall use below.

\subsection{Generalizing diagonal S-matrices}

Here we follow \cite{CF88} and propose a quite simple principle which
introduces an infinite number of unstable particles into the spectrum. We
note first, that in general many scattering matrices factorize in the
following form%
\begin{equation}
S_{ab}(\theta )=S_{ab}^{\min }(\theta )S_{ab}^{\text{CDD}}(\theta )\,.
\end{equation}%
Here $S_{ab}^{\min }(\theta )$ denotes the so-called minimal S-matrix which
satisfies the consistency relations i)--vii) of section 2. The CDD-factor 
\cite{CDD}, only satisfies i)-vi) and has its poles in the sheet $-\pi \leq 
\func{Im}\theta \leq 0$, which is the \textquotedblleft physical
one\textquotedblright\ for resonance states. We note now that the minimal
part is of the general form%
\begin{equation}
S_{ab}(\theta )=\prod_{x\in \mathcal{A}}\{x\}_{\theta }^{\sigma }\,,
\label{no}
\end{equation}%
with $\mathcal{A}$ being a finite set specific to each theory. Then we may
define a new $S$-matrix 
\begin{equation}
\hat{S}_{ab}(\theta )=\prod_{x\in \mathcal{A}}\{x\}_{\theta }^{\sigma
}\{x\}_{\theta ,\ell }^{\sigma }  \label{ns}
\end{equation}%
and note that the additional factor in (\ref{ns}) is just of CDD-type.
Therefore (\ref{ns}) constitutes a solution to the consistency relations
i)--vii) of section 2, and thus a strong candidate for a scattering matrix
of a proper quantum field theory. Note that wheras (\ref{no}) was a finite
product of hyperbolic functions, the new proposal (\ref{ns}) contains,
according to the identity (\ref{gen}) in addition elliptic functions, which
lead to the desired spectrum of infinitely many unstable particles according
to the principles outlined in section 2.

\subsection{Non-diagonal S-matrices}

We discuss now the elliptic sine-Gordon model, which may be related to the
continuum limit of the eight-vertex model. The (anti)-soliton sector was
studied many years ago in \cite{Z4}. In \cite{Breath} we demonstrated that
it is possible to associate a consistent breather sector to this model. Let
us recall the argument by recalling the Zamolodchikov algebra for the
soliton sector 
\begin{eqnarray}
Z(\theta _{1})Z(\theta _{2}) &=&a(\theta _{12})Z(\theta _{2})Z(\theta
_{1})+d(\theta _{12})\bar{Z}(\theta _{2})\bar{Z}(\theta _{1})~,  \label{ZZ1}
\\
Z(\theta _{1})\bar{Z}(\theta _{2}) &=&b(\theta _{12})\bar{Z}(\theta
_{2})Z(\theta _{1})+c(\theta _{12})Z(\theta _{2})\bar{Z}(\theta _{1})~.
\label{ZZ2}
\end{eqnarray}%
In comparison with the more extensively studied sine-Gordon model the
difference is the occurrence of the amplitude $d$ in (\ref{ZZ1}), i.e.~the
possibility that two solitons change into two anti-solitons and vice versa.
Invoking the consistency equations i)-v) one finds \cite{Z4,Breath}

\begin{eqnarray}
\quad \quad a(\theta ) &=&\Phi (\theta )\prod\limits_{k=0}^{\infty }\left( 
\frac{\Gamma _{\hat{q}^{2}}[-\hat{\theta}-\frac{1+2k}{2}\lambda ]\Gamma _{%
\hat{q}^{2}}[1-\hat{\theta}-\frac{1+2k}{2}\lambda ]}{\Gamma _{\hat{q}^{2}}[%
\hat{\theta}-\frac{1+2k}{2}\lambda ]\Gamma _{\hat{q}^{2}}[1+\hat{\theta}-%
\frac{1+2k}{2}\lambda ]}\right.  \label{a} \\
&&\qquad \qquad \times \left. \frac{\Gamma _{\hat{q}^{2}}[\hat{\theta}%
-(k+1)\lambda ]\Gamma _{\hat{q}^{2}}[1+\hat{\theta}-k\lambda ]}{\Gamma _{%
\hat{q}^{2}}[-\hat{\theta}-(k+1)\lambda ]\Gamma _{\hat{q}^{2}}[1-\hat{\theta}%
-k\lambda ]}\right)  \notag \\
b(\theta ) &=&-\frac{\func{sn}(i\theta /\nu )}{\func{sn}(i\theta /\nu +\pi
/\nu )}a(\theta ),  \label{b} \\
c(\theta ) &=&\frac{\func{sn}(\pi /\nu )}{\func{sn}(i\theta /\nu +\pi /\nu )}%
a(\theta ), \\
d(\theta ) &=&-\sqrt{\ell }\func{sn}(i\theta /\nu )\func{sn}(\pi /\nu
)a(\theta ),  \label{d} \\
\Phi (\theta ) &=&\frac{\Gamma _{\hat{q}}[1+\frac{\tau }{2}]\Gamma _{\hat{q}%
}[-\frac{\tau }{2}]\Gamma _{\hat{q}}[1-\hat{\theta}+\frac{\lambda }{2}+\frac{%
\tau }{2}]\Gamma _{\hat{q}}[\hat{\theta}-\frac{\lambda }{2}-\frac{\tau }{2}]%
}{\Gamma _{\hat{q}}[1+\hat{\theta}+\frac{\tau }{2}]\Gamma _{\hat{q}}[-\hat{%
\theta}-\frac{\tau }{2}]\Gamma _{\hat{q}}[1+\frac{\lambda }{2}+\frac{\tau }{2%
}]\Gamma _{\hat{q}}[-\frac{\lambda }{2}-\frac{\tau }{2}]}~.
\end{eqnarray}%
Here we used $\lambda =-\pi /K_{\ell }\nu $, $\hat{\theta}=i\theta /2K_{\ell
}\nu $ with $\nu \in \mathbb{R}$ being the coupling constant of the
model.With regard to property vii), it is clear that it is important to
analyse the singularity structure of the amplitudes (\ref{a})-(\ref{d}) to
judge whether there exists a breather sector. For this we appeal to the
relations (\ref{pG}), (\ref{zJ}) and (\ref{pJ}) and find the following pole
structure inside the physical sheet 
\begin{equation*}
\begin{array}{ll}
\theta _{a_{1},p}^{nm}=2m\nu K_{1-\ell }+i2n\nu K_{\ell }, & \theta
_{a_{2},p}^{nm}=(2m+1)\nu K_{1-\ell }+i(\pi -2n\nu K_{\ell }), \\ 
\theta _{b_{1},p}^{lm}=2m\nu K_{1-\ell }+i(\pi -2l\nu K_{\ell }),\quad & 
\theta _{b_{2},p}^{lm}=(2m+1)\nu K_{1-\ell }+i2l\nu K_{\ell }, \\ 
\theta _{c_{1},p}^{lm}=2m\nu K_{1-\ell }+i2l\nu K_{\ell }, & \theta
_{c_{2},p}^{lm}=2m\nu K_{1-\ell }+i(\pi -2l\nu K_{\ell }), \\ 
\theta _{d_{1},p}^{lm}=(2m+1)\nu K_{1-\ell }+i2l\nu K_{\ell }, & \theta
_{d_{2},p}^{lm}=(2m+1)\nu K_{1-\ell }+i(\pi -2n\nu K_{\ell }).%
\end{array}%
\end{equation*}

\noindent We took $l,m\in \mathbb{Z},n\in \mathbb{N}$ and associated always
two sets of poles $\theta _{a_{1},p}^{nm}$ and $\theta _{a_{2},p}^{nm}$ to $%
a(\theta )$, $\theta _{b_{1},p}^{nm}$ and $\theta _{b_{2},p}^{nm}$ to $%
b(\theta )$ etc. One readily sees from this that if one restricts the
parameter $\nu \geq \pi /2K_{\ell }$ all poles move out of the physical
sheet into the non-physical one, where they can be interpreted in principle
as unstable particles. This was already stated in \cite{Z4}, where the
choice $\nu \geq \pi /2K_{\ell }$ was made in order to avoid the occurrence
of non-physical states. This is clear from our discussion of property vii)
in section 2, as we would have poles in the physical sheet beyond the
imaginary axis, which when interpreted with the Breit-Wigner formula leave
the choice that either $m_{\bar{c}}<0$ or $\Gamma _{\bar{c}}<0$, i.e. we
either violate causality or we have Tachyons. The restriction on the
parameters makes the model somewhat unattractive as this limitation
eliminates the analogue of the entire breather sector which is present in
the sine-Gordon model, such that also in the trigonometric limit one only
obtains the soliton-antisoliton sector of that model, instead of a theory
with a richer particle content. For this reason we relax here the
restriction on $\nu $ and note that the poles 
\begin{equation}
\theta _{b_{1},p}^{n0}=\theta _{c_{2},p}^{n0}\qquad \quad ~~\text{for~\ }%
0<n<n_{\max }=[\pi /2\nu K_{\ell }],n\in \mathbb{N}  \label{bp}
\end{equation}%
are located on the imaginary axis inside the physical sheet and are
therefore candidates for the analogue of the n$^{th}$-breather bound states
in the sine-Gordon model. We indicate here the integer part of $x$ by $[x]$.
In other words, there are at most $n_{\max }-1$ breathers for fixed $\nu $
and $\ell $. The price one pays for the occurrence of these new particles in
the elliptic sine-Gordon model is that one unavoidably also introduces
additional Tachyons into the model as the poles always emerge in
\textquotedblleft strings\textquotedblright . It remains to be established
whether the poles (\ref{bp}) may really be associated to a breather type
behaviour.

Let us now see if the poles on the imaginary axis inside the physical sheet
can be associated consistently with breathers. We proceed similarly as for
the sine-Gordon model \cite{S13}, even though in the latter approach the
following ansatz is inspired by the classical theory and here we do not have
a classical counterpart. We define the auxiliary state \ 
\begin{equation}
Z_{n}(\theta _{1},\theta _{2}):=\frac{1}{\sqrt{2}}\left[ Z(\theta _{1})\bar{Z%
}(\theta _{2})+(-1)^{n}\bar{Z}(\theta _{1})Z(\theta _{2})\right] ~.
\label{breather}
\end{equation}%
This state has properties of the classical sine-Gordon breather being
chargeless and having parity $(-1)^{n}$. Choosing thereafter the rapidities
such that the state (\ref{breather}) is on-shell, we can speak of a breather
bound state 
\begin{equation}
\lim_{(p_{1}+p_{2})^{2}\rightarrow m_{b_{n}}^{2}}Z_{n}(\theta _{1},\theta
_{2})\equiv \lim_{\theta _{12}\rightarrow \theta +\theta
_{12}^{b_{n}}}Z_{n}(\theta _{1},\theta _{2})=Z_{n}(\theta )~.
\label{onshell}
\end{equation}%
Here $\theta _{12}^{b_{n}}$ is the fusing angle related to the poles in the
soliton-antisoliton scattering amplitudes. We compute now with the help of (%
\ref{ZZ1}) and (\ref{ZZ2}) the exchange relation 
\begin{equation}
Z_{n}(\theta _{1})Z(\theta _{2})=S_{b_{n}s}(\theta _{12})Z(\theta
_{2})Z_{n}(\theta _{1})~,  \label{bs}
\end{equation}%
where 
\begin{equation}
S_{b_{n}s}(\theta )=\frac{\func{sn}(\frac{i\theta }{\nu }-\frac{\pi }{2\nu }%
+nK_{\ell })}{\func{sn}(\frac{i\theta }{\nu }+\frac{\pi }{2\nu }+nK_{\ell })}%
[\ell {\func{sn}}^{2}\frac{\pi }{\nu }{\func{sn}}^{2}\left( \frac{i\theta }{%
\nu }+\frac{\pi }{2\nu }+nK_{\ell }\right) -1]\bar{a}\quad  \label{sbs}
\end{equation}%
and 
\begin{eqnarray*}
&&\bar{a}=\frac{\Gamma _{\hat{q}^{2}}[1+\hat{\theta}+\frac{\lambda }{4}-%
\frac{n}{2}]\Gamma _{\hat{q}^{2}}[-\hat{\theta}-\frac{\lambda }{4}-\frac{n}{2%
}]\Gamma _{\hat{q}^{2}}[-\hat{\theta}+\frac{\lambda }{4}+\frac{n}{2}]\Gamma
_{\hat{q}^{2}}[\hat{\theta}+\frac{\lambda }{4}-\frac{n}{2}]}{\Gamma _{\hat{q}%
^{2}}[1-\hat{\theta}+\frac{\lambda }{4}-\frac{n}{2}]\Gamma _{\hat{q}^{2}}[%
\hat{\theta}-\frac{\lambda }{4}-\frac{n}{2}]\Gamma _{\hat{q}^{2}}[\hat{\theta%
}+\frac{\lambda }{4}+\frac{n}{2}]\Gamma _{\hat{q}^{2}}[-\hat{\theta}+\frac{%
\lambda }{4}-\frac{n}{2}]} \\
&&\times \Phi _{13}\Phi _{23}\prod\limits_{l=1}^{n-1}\frac{[\hat{\theta}-%
\frac{n}{2}+\frac{\lambda }{4}-k\lambda +l]_{\hat{q}^{2}}^{2}[-\hat{\theta}+%
\frac{n}{2}-\frac{\lambda }{4}-k\lambda -l]_{\hat{q}^{2}}^{2}}{[-\hat{\theta}%
-\frac{n}{2}+\frac{\lambda }{4}-k\lambda +l]_{\hat{q}^{2}}^{2}[\hat{\theta}+%
\frac{n}{2}-\frac{\lambda }{4}-k\lambda -l]_{\hat{q}^{2}}^{2}} \\
&&\times \prod\limits_{k=0}^{\infty }\frac{[\hat{\theta}-\frac{n}{2}+\frac{%
\lambda }{4}-k\lambda ]_{\hat{q}^{2}}[-\hat{\theta}+\frac{n}{2}-\frac{%
\lambda }{4}-k\lambda ]_{\hat{q}^{2}}}{[-\hat{\theta}-\frac{n}{2}+\frac{%
\lambda }{4}-k\lambda ]_{\hat{q}^{2}}[\hat{\theta}+\frac{n}{2}-\frac{\lambda 
}{4}-k\lambda ]_{\hat{q}^{2}}} \\
&&\times \prod\limits_{k=0}^{\infty }\frac{[\hat{\theta}+\frac{n}{2}+\frac{%
\lambda }{4}-k\lambda ]_{\hat{q}^{2}}[-\hat{\theta}-\frac{n}{2}-\frac{%
\lambda }{4}-k\lambda ]_{\hat{q}^{2}}}{[-\hat{\theta}+\frac{n}{2}+\frac{%
\lambda }{4}-k\lambda ]_{\hat{q}^{2}}[\hat{\theta}-\frac{n}{2}-\frac{\lambda 
}{4}-k\lambda ]_{\hat{q}^{2}}}
\end{eqnarray*}%
Where $\Phi _{ij}=\Phi (\theta _{ij})$ with $\theta _{ij}$ being the
difference of the on-shell rapidities. What is remarkable here and can not
be anticipated a priori, is that all off-diagonal terms vanish, thus as (\ref%
{bs}) expresses in the soliton breather scattering there is no
backscattering. Similarly, but more lengthy, we compute the scattering
amplitude between the n$^{th}$-breather \ and m$^{th}$-breather 
\begin{equation}
Z_{n}(\theta _{1})Z_{m}(\theta _{2})=S_{b_{n}b_{m}}(\theta
_{12})Z_{m}(\theta _{2})Z_{n}(\theta _{1})
\end{equation}%
where

\begin{equation}
S_{b_{n}b_{m}}(\theta )=\left[ 1-\ell {\func{sn}}^{2}\frac{\pi }{\nu }{\func{%
sn}}^{2}\left( \frac{i\theta }{\nu }+(n+m)K_{\ell }+\frac{\pi }{\nu }\right) %
\right] \qquad \qquad \qquad \qquad  \label{Sbb2}
\end{equation}%
\begin{equation}
\qquad \qquad \times \left[ 1-\ell {\func{sn}}^{2}\frac{\pi }{\nu }{\func{sn}%
}^{2}\left( \frac{i\theta }{\nu }+(n+m)K_{\ell }\right) \right] \frac{\func{%
sn}(i\theta /\nu -\pi /\nu +(n+m)K_{\ell })}{\func{sn}(i\theta /\nu +\pi
/\nu +(n+m)K_{\ell })}\tilde{a}  \notag
\end{equation}%
and 
\begin{eqnarray}
\tilde{a} &{=}&{\Phi }_{13}{\Phi }_{14}{\Phi }_{23}{\Phi }_{24}\frac{\Gamma
_{\hat{q}^{2}}(1+{\frac{m}{2}}+{\frac{n}{2}}+\hat{\theta}+{\frac{\lambda }{2}%
})\,\Gamma _{\hat{q}^{2}}({\frac{-m}{2}}-{\frac{n}{2}}-\hat{\theta}-{\frac{%
\lambda }{2}})}{\Gamma _{\hat{q}^{2}}(1+{\frac{m}{2}}+{\frac{n}{2}}-\hat{%
\theta}+{\frac{\lambda }{2}})\Gamma _{\hat{q}^{2}}({\frac{-m}{2}}-{\frac{n}{2%
}}+\hat{\theta}-{\frac{\,\lambda }{2}})}  \notag \\
&&\times \prod\limits_{k=1}^{\infty }\prod\limits_{l=1}^{n-1}\frac{[{\frac{m%
}{2}}+{\frac{n}{2}-l}-\hat{\theta}-k\,\lambda +\lambda ]_{\hat{q}^{2}}[{%
\frac{-m}{2}}-{\frac{n}{2}+l}+\hat{\theta}-k\,\lambda ]_{\hat{q}^{2}}}{[{%
\frac{m}{2}}+{\frac{n}{2}-l}+\hat{\theta}-k\,\lambda +\lambda ]_{\hat{q}%
^{2}}[{\frac{-m}{2}}-{\frac{n}{2}+l}-\hat{\theta}-k\,\lambda ]_{\hat{q}^{2}}}
\notag \\
&&\times \prod\limits_{k=0}^{\infty }\prod\limits_{l=1}^{n-1}\frac{[{\frac{m%
}{2}}+{\frac{n}{2}-l}+\hat{\theta}-{\frac{\lambda }{2}}-k\,\lambda ]_{\hat{q}%
^{2}}[-{\frac{m}{2}}-{\frac{n}{2}+l}-\hat{\theta}-{\frac{\lambda }{2}}%
-k\,\lambda ]_{\hat{q}^{2}}}{[{\frac{m}{2}}+{\frac{n}{2}-l}-\hat{\theta}-{%
\frac{\lambda }{2}}-k\,\lambda ]_{\hat{q}^{2}}[-{\frac{m}{2}}-{\frac{n}{2}+l}%
+\hat{\theta}-{\frac{\lambda }{2}}-k\,\lambda ]_{\hat{q}^{2}}}  \label{aaaaa}
\\
&&\times \prod\limits_{k=1}^{\infty }\prod\limits_{l=0}^{m-1}\frac{[{\frac{m%
}{2}}+{\frac{n}{2}-l}-\hat{\theta}-k\,\lambda +\lambda ]_{\hat{q}^{2}}[-{%
\frac{m}{2}}-{\frac{n}{2}+l}+\hat{\theta}-k\,\lambda ]_{\hat{q}^{2}}}{[{%
\frac{m}{2}}+{\frac{n}{2}-l}+\hat{\theta}-k\,\lambda +\lambda ]_{\hat{q}%
^{2}}[-{\frac{m}{2}}-{\frac{n}{2}+l}-\hat{\theta}-k\,\lambda ]_{\hat{q}^{2}}}
\notag \\
&&\times \prod\limits_{k=0}^{\infty }\prod\limits_{l=0}^{m-1}\frac{[{\frac{m%
}{2}}+{\frac{n}{2}-l}+\hat{\theta}-{\frac{\lambda }{2}}-k\,\lambda ]_{\hat{q}%
^{2}}[-{\frac{m}{2}}-{\frac{n}{2}+l}-\hat{\theta}-{\frac{\lambda }{2}}%
-k\,\lambda ]_{\hat{q}^{2}}}{[{\frac{m}{2}}+{\frac{n}{2}-l}-\hat{\theta}-{%
\frac{\lambda }{2}}-k\,\lambda ]_{\hat{q}^{2}}[-{\frac{m}{2}}-{\frac{n}{2}+l}%
+\hat{\theta}-{\frac{\lambda }{2}}-k\,\lambda ]_{\hat{q}^{2}}}.  \notag
\end{eqnarray}%
The latter expression (\ref{aaaaa}) is tailored to make contact to the
expressions in the literature corresponding to the trigonometric limit. Also
for this amplitude the backscattering is zero.

The matrix $S_{b_{n}b_{m}}(\theta )$ also exhibits several types of poles.
a) simple and double poles inside the physical sheet beyond the imaginary
axis, b) double poles located on the imaginary axis, c) simple poles in the
non-physical sheet and d one simple pole on the imaginary axis inside the
physical sheet at $\theta =\theta _{b}=i\nu (n+m)K_{\ell }$ which is related
to the fusing process of two breathers $b_{n}+b_{m}\rightarrow b_{n+m}$. To
be really sure that this pole admits such an interpretation, we have to
establish according to (\ref{sing}) that the imaginary part of the residue
is strictly positive, i.e. 
\begin{equation}
-i\lim_{\theta \rightarrow \theta _{b}}(\theta -\theta
_{b})S_{b_{n}b_{m}}(\theta )>0~.  \label{sheap}
\end{equation}%
The explicit computation shows that this is indeed the case, see \cite%
{Breath}.

Furthermore, It is very interesting to check if also (\ref{boot}) is
satisfied for the fusing process $b_{n}+b_{m}\rightarrow b_{n+m}$. For
consistency, all amplitudes have to satisfy the bootstrap equations 
\begin{equation}
S_{lb_{n+m}}(\theta )=S_{lb_{n}}(\theta +i\nu mK_{\ell })S_{lb_{m}}(\theta
-i\nu nK_{\ell })  \; ,
\label{bboot}
\end{equation}%
for  $ l\in \{b_{k},s,\bar{s}\}~;k,m+n<n_{\max }$. Indeed, we verify with some algebra that (\ref{bboot}) holds for the above
amplitudes (\ref{sbs}) and (\ref{Sbb2}).

Finally, we carry out various limits. Our formulation in terms of q-deformed
quantities and elliptic functions is useful to make this task fairly easy.
We state our results here only schematically and refer the reader for
details to \cite{Breath}. We find 
\begin{equation*}
\begin{array}{ccc}
\fbox{elliptic sine-Gordon} & \overset{1/\nu \rightarrow 2nK_{\ell }/\pi
+2imK_{1-\ell }/\pi }{-------\longrightarrow } & \fbox{elliptic $%
D_{n+1}^{(1)}$-ATFT} \\ 
\begin{array}{c}
| \\ 
| \\ 
\ell \rightarrow 0 \\ 
| \\ 
\downarrow%
\end{array}
& 
\begin{array}{ccc}
&  & \swarrow \\ 
& 
\begin{array}{c}
m\neq 0,\ell \rightarrow 0 \\ 
\downarrow \\ 
\fbox{free theory} \\ 
\uparrow \\ 
1/\nu \rightarrow i\infty%
\end{array}
&  \\ 
\nearrow &  & 
\end{array}
& 
\begin{array}{c}
| \\ 
| \\ 
m=0,\ell \rightarrow 0 \\ 
| \\ 
\downarrow%
\end{array}
\\ 
\fbox{sine-Gordon} & \underset{1/\nu \rightarrow n}{-------\longrightarrow }
& \fbox{minimal $D_{n+1}^{(1)}$-ATFT}%
\end{array}%
\end{equation*}%
Thus we can view the elliptic sine-Gordon model as a master theory for
several other models. In the limit $\ell \rightarrow 0$ we recover now all
sectors, including the breathers, of the sine-Gordon model. The diagonal
limit $1/\nu \rightarrow 2nK_{\ell }/\pi +2imK_{1-\ell }/\pi $ is
interesting as it yields a new type of theory, which we refer to as elliptic 
$SO(2n+2)\equiv D_{n+1}^{(1)}$-affine Toda field theory (ATFT). To coin this
name for these theories seems natural as in the trigonometric limit we
obtain from it the ordinary minimal $D_{n+1}^{(1)}$-ATFT.

\section{Conclusions}

We reviewed the general analytical scattering theory related to integrable
quantum field theories in 1+1 space-time dimensions. We made a proposal for
a construction principle of an S-matrix like object which describes the
scattering between two unstable particles or an unstable particle and a
stable one. We tested this proposal with various examples and found a
remarkable agreement with the outcome of the thermodynamic Bethe ansatz in
what concerns the particle content and the RG flow of the theories. We
described the general Lie algebraic structure of theories with unstable
particles and propose a decoupling rule which predicts the RG flow when some
of the parameters in the theory become very large. Alternatively, we tested
these analytical prediction with the TBA. Finally, we discussed how one can
construct theories with and without backscattering which contain an infinite
number of unstable particles.

\subsection*{Acknowledgment}

We are grateful to the Deutsche
Forschungsgemeinschaft (Sfb288), for financial support. This work is supported by the EU 
network EUCLID, \emph{Integrable models and
applications: from strings to condensed matter}, HPRN-CT-2002-00325.

\end{document}